\documentclass[manuscript]{aastex631} %linenumbers,

\accepted{\today}
\submitjournal{PSJ}

\shorttitle{Venus dayside observation campaign 2020}
\shortauthors{Lee et al.}
\graphicspath{{./}}

\begin{document}

\title{Reflectivity of Venus' dayside disk during the 2020 observation campaign: outcomes and future perspectives}

\correspondingauthor{Yeon Joo Lee}
\email{yeonjoolee@ibs.re.kr}%yjlee.msg@gmail.com}

\author[0000-0002-4571-0669]{Yeon Joo Lee}
\affiliation{DLR Institute of Planetary Research, Berlin, Germany}
\affiliation{Institute for Basic Science, Daejeon, South Korea}

\author[0000-0003-1756-4825]{Antonio Garc\'ia~Mu\~noz}
\affiliation{AIM, CEA, CNRS, Universit\'e Paris-Saclay, Universit\'e de Paris, Gif-sur-Yvette, France}

\author[0000-0001-6468-6812]{Atsushi Yamazaki}
\affiliation{Institute of Space and Astronautical Science (ISAS/JAXA), Sagamihara, Japan}
\affiliation{Graduate School of Science, University of Tokyo, Tokyo, Japan}

\author[0000-0001-5376-2242]{Eric Qu\'emerais}
\affiliation{LATMOS-OVSQ, Universit\'e Versailles Saint-Quentin, Guyancourt, France}

\author{Stefano Mottola}
\affiliation{DLR Institute of Planetary Research, Berlin, Germany}

\author[0000-0003-3997-3363]{Stephan Hellmich}
\affiliation{DLR Institute of Planetary Research, Berlin, Germany}
\affiliation{Laboratory of astrophysics, \'Ecole Polytechnique F\'ed\'erale de Lausanne (EPFL), Observatoire de Sauverny, 1290 Versoix, Switzerland}

\author{Thomas Granzer}
\affiliation{Leibniz-Institute for Astrophysics Potsdam (AIP), Potsdam, Germany}

\author{Gilles Bergond}
\affiliation{CAHA, Almeria, Spain}

\author[0000-0003-2451-739X]{Martin Roth}
\affiliation{Leibniz-Institute for Astrophysics Potsdam (AIP), Potsdam, Germany}

\author[0000-0002-7452-1496]{Eulalia Gallego-Cano}
\affiliation{IAA Granada, Spain}

\author{Jean-Yves Chaufray}
\affiliation{LATMOS-OVSQ, Universit\'e Versailles Saint-Quentin, Guyancourt, France}

\author{Rozenn Robidel}
\affiliation{LATMOS-OVSQ, Universit\'e Versailles Saint-Quentin, Guyancourt, France}

\author{Go Murakami}
\affiliation{Institute of Space and Astronautical Science (ISAS/JAXA), Sagamihara, Japan}

\author[0000-0001-9704-6993]{Kei Masunaga}
\affiliation{Institute of Space and Astronautical Science (ISAS/JAXA), Sagamihara, Japan}

\author{Murat Kaplan}
\affiliation{Akdeniz Univ., Antalya, Turkey}

\author{Orhan Erece}
\affiliation{Akdeniz Univ., Antalya, Turkey}

\author[0000-0003-0169-123X]{Ricardo Hueso}
\affiliation{Dpt. F\'sica Aplicada, Escuela de Ingenier\'a de Bilbao, Universidad del Pa\'s Vasco UPV/EHU, Bilbao, Spain}

\author{Petr Kab\'ath}
\affiliation{Astronomical Institute AS CR, Ondrejov, Czech Republic}

\author{Magdal\'{e}na \v{S}pokov\'{a}}
\affiliation{Astronomical Institute AS CR, Ondrejov, Czech Republic}
\affiliation{Masaryk University, Department of theoretical physics and astrophysics, Kotl\'{a}\v{r}sk\'{a} 2, 611 37, Brno, Czech Republic}

\author[0000-0001-7234-7634]{Agust\'in S\'anchez-Lavega}
\affiliation{Dpt. F\'sica Aplicada, Escuela de Ingenier\'a de Bilbao, Universidad del Pa\'s Vasco UPV/EHU, Bilbao, Spain}

\author{Myung-Jin Kim}
\affiliation{Korea Astronomy and Space Science Institute (KASI), Daejeon, South Korea}

\author[0000-0002-9903-4053]{Valeria Mangano}
\affiliation{INAF-IAPS (Institute for Astrophysics and Planetology from Space), Rome, Italy}

\author{Kandis-Lea Jessup}
\affiliation{Southwest Research Institute, Boulder, CO, USA}

\author{Thomas Widemann}
\affiliation{Observatoire de Paris-PSL \& Universit\'e Paris-Saclay, LESIA - UMR CNRS, Meudon, France}

\author{Ko-ichiro Sugiyama}
\affiliation{Matsue National College of Technology, Matsue, Japan}

\author[0000-0002-3058-0689]{Shigeto Watanabe}
\affiliation{Space Information Center, Hokkaido Information University, Ebetsu, Japan}

\author[0000-0003-0726-6592]{Manabu Yamada}
\affiliation{Planetary Exploration Research Center (PERC), Narashino, Japan}

\author[0000-0001-9071-5808]{Takehiko Satoh}
\affiliation{Institute of Space and Astronautical Science (ISAS/JAXA), Sagamihara, Japan}

\author{Masato Nakamura}
\affiliation{Institute of Space and Astronautical Science (ISAS/JAXA), Sagamihara, Japan}

\author[0000-0001-8543-6556]{Masataka Imai}
\affiliation{Kyoto Sangyo University, Kyoto, Japan}

\author[0000-0001-6653-5487]{Juan Cabrera}
\affiliation{DLR Institute of Planetary Research, Berlin, Germany}

\begin{abstract}
We performed a unique Venus observation campaign to measure the disk brightness of Venus over a broad range of wavelengths in August and September 2020. The primary goal of the campaign is to investigate the absorption properties of the unknown absorber in the clouds. The secondary goal is to extract a disk mean SO$_2$ gas abundance, whose absorption spectral feature is entangled with that of the unknown absorber at the ultraviolet (UV) wavelengths. A total of 3 spacecraft and 6 ground-based telescopes participated in this campaign, covering the 52 to 1700~nm wavelength range. After careful evaluation of the observational data, we focused on the data sets acquired by 4 facilities. We accomplished our primary goal by analyzing the reflectivity spectrum of the Venus disk over the 283-800 nm wavelengths. Considerable absorption is present in the 350-450 nm range, for which we retrieved the corresponding optical depth by the unknown absorber. The result shows a consistent wavelength dependence of the relative optical depth with that at low latitudes during the Venus flyby by MESSENGER in 2007 \citep{Perezhoyos18}, which was expected because the overall disk reflectivity is dominated by low latitudes. Last, we summarize the experience obtained during this first campaign that should enable us to accomplish our second goal in future campaigns.

\end{abstract}

\section{Introduction} \label{sec:intro}
As the third brightest object in the sky after the Sun and the Moon, the scientific observations of Venus started early. A century ago, ground-based observations discovered the presence of dark patches in UV images of the planet \citep{Wright27,Ross28}. The chemical that produces the dark patches on the planet is characterized by broad absorption that extends from the UV to the visible wavelengths. The identity of such chemical remains elusive and the substance is still called the ``unknown absorber''\citep{Barker75,Pollack80b,Zasova81,Mills07,Titov18}. Recent studies suggested that the unknown absorber may be OSSO or S$_2$O which explains the observed UV spectrum \citep{Perezhoyos18}. According to photochemical model calculations \citep{Krasnopolsky18} and glory observation analysis \citep{Petrova18}, the unknown absober could also be iron chloride. There are more candidates, such as S$_{\rm{x}}$, Cl$_2$, SCl$_2$, etc. \citep{Mills07}. Recently, iron-bearing micro-organisms have also been proposed \citep{Limaye18}.

The absorption spectrum of the unknown absorber was reported to have its maximum at 340~nm with a FWHM of 140~nm, according to the MESSENGER/MASCS data \citep{Perezhoyos18}. But considering the limited spectral range of the MESSENGER/MASCS data: 300$-$1500~nm, the spectral properties of the unknown absorber at $\lambda$$<$300~nm was not accessed, remaining undefined. Spectral data at such short wavelengths were acquired by the SPICAV spectrometer on board Venus Express, covering the 170$-$320~nm range by its UV channel. In order to explain the data taken by SPICAV's UV channel, \citet{Marcq11,Marcq19} postulated the presence of an unknown absorber in the form of a cloud aerosol, in addition to pure sulfuric acid aerosol. The putative absorber would explain the absorption shortward of 300~nm. These previous studies suggest that the unknown absorber remains effective at wavelengths from $\sim$200~nm \citep{Marcq19} to $\sim$600~nm \citep{Perezhoyos18}. These observations were done at different times and with different viewing geometries, so their data cannot be directly combined to understand the spectral properties of the unknown absorber over the entire UV$-$visible wavelength range. To elucidate such properties, it is clear that additional observations should be made over a broader range of wavelengths, such as those done by the STIS spectrometer aboard Hubble Space Telescope over the 200$-$600~nm \citep{Jessup19}.

The UV observations are also useful to retrieve abundances of trace gases near the cloud top level. For example, SO$_2$ bands are located near 215 and 280~nm, SO band near 215~nm, and O$_3$ band near 250~nm \citep{Esposito88,Na90,Belyaev12,Jessup15,Marcq19_O3,Marcq19}. Their abundances and variations are important to understand photochemical processes in the atmosphere \citep{Mills07,Titov18}, including their interaction with the unknown absorber \citep{Lee15a,Lee19,Marcq13,Marcq19}. However, without high spectral resolution, the interpretation is complicated by the overlap of the bands and by the absorption of the unknown absorber. A further complication would be represented by the presence of an additional species, H$_2$S near the cloud top level, as suggested by \citet{Bierson20}. This contribution, not considered in previous studies \citep{Na90,Belyaev12,Jessup15}, is characterized by a UV band near 215 nm that overlaps those of the SO and SO$_2$ gases.

Significant temporal variations of the unknown absorber and SO$_2$ gas abundance have been reported over both short- and long-term periods \citep{DelGenio82,DelGenio90,Esposito88,Imai19,Lee15a,Lee19,Lee20,Marcq13,Marcq19}. In terms of disk-integrated UV brightness, short-term variations indicate the presence of global-scale atmospheric waves with a 4$-$5~days periodicity \citep{DelGenio82,Lee20}, whose amplitudes are changing with time \citep{DelGenio90,Imai19,Lee20}. Changes of disk-integrated UV brightness over time scales of decades can impact the solar energy deposition in the atmosphere because almost half of solar heating at the cloud top atmosphere is caused by the unknown absorber \citep{Crisp86,Lee15b}. The latter can lead to considerable changes in global scale circulation and zonal wind speeds \citep{Lee19}. Intriguingly, the UV brightness variations are correlated with the SO$_2$ gas abundance near the cloud top level \citep{Lee15a,Lee19,Marcq19}. That connection is key to understanding the photochemical processes that affect cloud formation \citep{Mills07} and the impact of possible volcanic outgasing on the atmosphere. We need further data to investigate the relationship between the sulfur-related gaseous abundance and the unknown absorber. That is the main motivation for the Venus dayside observation campaign that we performed in 2020.

As our campaign measures disk-integrated spectral brightness, the results will be useful for comparison with spatially unresolved data acquired by future exoplanet imaging investigations. For example, we now know that measuring the planet's brightness at more than one phase angle can be a valuable strategy to identify Venus-like clouds at exoplanets, if they exist, by future direct imaging telescopes \citep{Carrion-Gonzalez20,Carrion-Gonzalez21}. In this manuscript we describe the campaign (Sect.~\ref{sec:obs}), explain the data reduction (Sect.~\ref{sec:data}), the atmospheric modeling (Sect.~\ref{sec:model}), and the data analysis (Sect.~\ref{sec:results}), and offer our lessons learned for the purpose of planning future campaigns (Sect.~\ref{sec:discussion}).

\section{Observations} \label{sec:obs}
\begin{figure*}
	\gridline{
		\fig{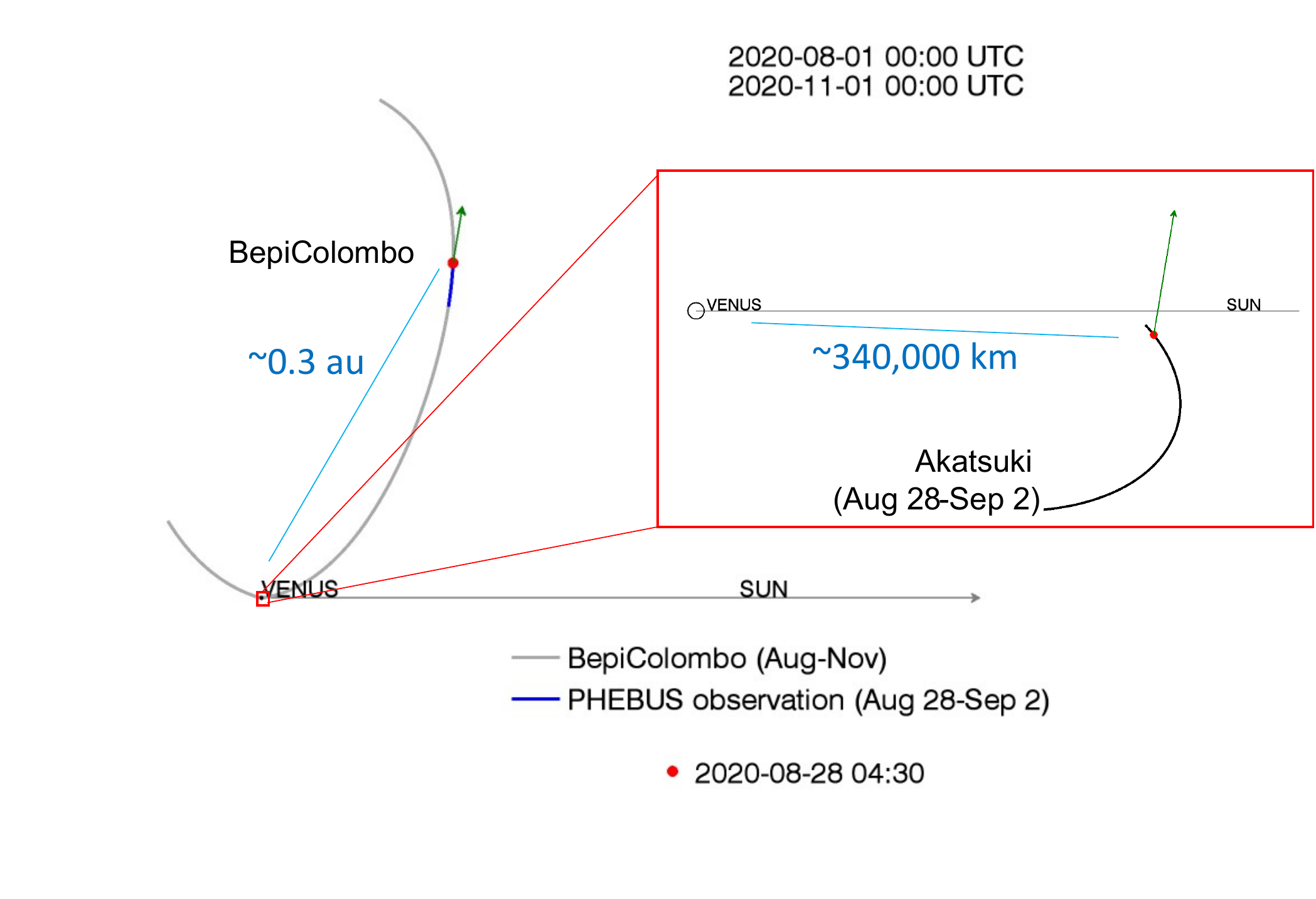}{0.8\textwidth}{(a)}
	}
	\gridline{
		\fig{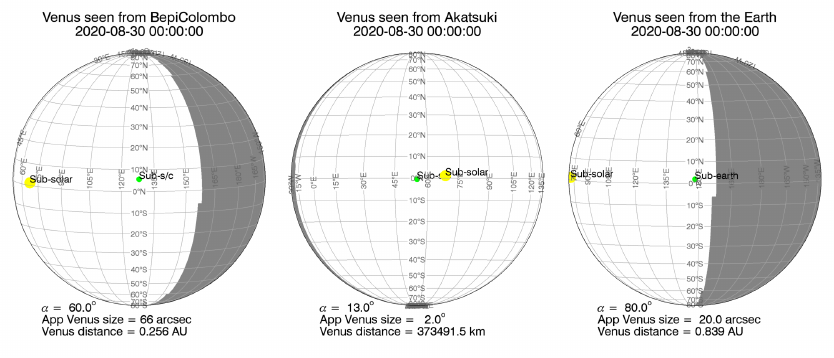}{0.8\textwidth}{(b)}
	}
	\caption{Overview of the campaign observations. (a) Relative locations of observation facilities from Venus. The green arrows indicate the direction towards Earth from the spacecraft. The gray curve of BepiColombo indicates the trajectory from Aug to Nov in 2020, and the blue curve highlights the location during the campaign period (Aug 28--Sep 2). Akatsuki's trajectory on Aug 28--Sep 2 is shown in the enlarged red box. The red dots are the locations of the spacecraft on Aug 28 at 04:30~UT. (b) Viewing geometries of Venus from Akatsuki, BepiColombo, and Earth on Aug 30. Day/night areas are indicated with white/gray areas over the disk. Solar phase angle ($\alpha$), apparent size of Venus, and the distance between the planet and observers are listed in the bottom. The sub-Solar and the sub-observer points are indicated with yellow and green dots, respectively. \label{fig:campaign_geometry}}
\end{figure*}

In August and September 2020 we performed the Venus dayside observation campaign from three locations in the Solar System: the Akatsuki Venus orbiter, the BepiColombo Mercury orbiter on its cruise phase towards Mercury, and the Earth (Earth-orbiting Hisaki spacecraft and ground-based telescopes) (Fig.~\ref{fig:campaign_geometry}a). JAXA's Venus orbiter Akatsuki operates from a highly elliptical equatorial orbit. The on-board UV camera (UVI) has monitored Venus since the orbit insertion in 2015 December \citep{Nakamura16}. ESA-JAXA's BepiColombo conducted faraway Venus observations from a distance of 0.3~au in the period of August 28$-$September 2, 2020 when Venus was within the Field-of-View (FOV) of the on-board UV spectrometer (PHEBUS) \citep{Mangano21}. While these two spacecraft were operating, ground-based telescopes were in a good condition to observe Venus for more than an hour right before sunrise. Three telescopes of the Calar Alto observatory (CAHA) joined the campaign and conducted the Venus observations: CAHA 1.23m DLR-MKIII CCD camera\footnote{\url{http://www.caha.es/CAHA/Instruments/IA123/DLR_Observation_guide_v1.11.pdf}}, CAHA 2.2m PlanetCam camera \citep{Mendikoa16}, and CAHA 3.5m Potsdam Multi-Aperture Spectrophotometer (PMAS) imaging-spectrometer \citep{Roth05}. T\"{U}B\.{I}TAK Naional Observatory's T100 CCD camera\footnote{\url{https://tug.tubitak.gov.tr/en/teleskoplar/t100-telescope}} and STELLA 1.2m telescope's Wide-Field STELLA Imaging Photometer (WiFSIP) \citep{Strassmeier10} acquired images, and Perek telescope's Ond\v{r}ejov Echelle Spectrograph (OES) \citep{Kabath20} acquired spectra. JAXA's Earth-orbiting Hisaki space telescope also obtained Venus data in the extreme ultraviolet (EUV) range with the EXCEED spectrometer \citep{Yoshikawa14}, which has been used to detect airglow of Venus \citep{Nara18}. The EUV data can help examine possible faint dayside reflection by the upper haze of Venus, thanks to their long exposure time over ten days. Table~\ref{tab:obs_summary} shows the complete list of facilities with observation dates and wavelength ranges.

The uniqueness of this campaign is the broad spectral coverage for observations of the Venus disk that extends from 52 to 1700~nm, and that cannot be acquired by a single instrument. We took advantage of the spectral overlap between the instruments, which could be used to combine individual spectral pieces of the brightness. For example, EXCEED and PHEBUS overlap at 145--148~nm; PHEBUS and UVI at 283~nm; UVI, ground-based U band and PMAS at 365~nm; ground-based B band and PMAS at 445~nm.

Half of the facilities acquired data of sufficient quality for scientific analysis, but not the others (Table~\ref{tab:obs_summary}). There were four problems for the latter. (1) The first problem was uncertainties in pointing as occurred during the data acquisition by PHEBUS and EXCEED. Narrow-slit spectrometers require high accuracy of spacecraft attitude control. The Venus observations by BepiColombo were in fact part of the performance tests on the cruise phase, and turned out that pointing accuracy was not always as good as planned. Hisaki gradually saw such control deteriorating with aging. Regardless of this problem, PHEBUS and EXCEED data could have been sufficient for relative spectral analysis. But PHEBUS data have an additional issue; its effective area turned out to be not well defined for scientific analysis (Sect.~\ref{subsec:phebus}). Consequently, the EXCEED data could not be used as the data comparison is not possible at the overlapping wavelengths (145--148~nm). Also spectral comparison of reflected daylight between EXCEED and PHEBUS may not be possible even in future, because PHEBUS' exposure time cannot be as long as EXCEED's. (2) The second problem was the photometric calibration of ground-based measurements, which required particular care. To define the telluric extinction coefficients accurately, CAHA1.23 DLR-MKIII camera frequently interspersed measurements of reference stars with those of Venus (Sect.~\ref{subsec:caha1.23}), and STELLA WiFSIP measured a reference star continuously until Venus rose sufficiently high to acquire data (Sect.~\ref{subsec:stella}). However, this was not the case for the other facilities, such as CAHA2.2 PlanetCam whose frequency of reference star observations turned out to be insufficient for photometric analysis under variable sky conditions. (3) The third problem was difficulties in defining an optimal aperture size in the CCD aperture photometry analysis. The TUG T100 data suffered from this problem, which may be exacerbated by the brightness of Venus. (4) For the last, Perek OES measurements are not used in this study, because its 2$''$ width slit is likely on the morning terminator (center of the 20$''$ diameter Venus disk).

Venus observations were conducted in three solar phase angle ($\alpha$) ranges as shown in Fig.~\ref{fig:campaign_geometry}b. Near the end of August, $\alpha$ was 60$^{\circ}$ for PHEBUS, 0--40$^{\circ}$ for UVI, and $\sim$80$^{\circ}$ from the Earth. In this manuscript we investigate the spectral features of the entire Venus disk at $\alpha=80^{\circ}$. To that end, we approximately corrected all the observations at other phase angles to form equivalent observations at  $\alpha=80^{\circ}$. In the future we plan to investigate the solar phase angle dependence of Venus' brightness \citep{Lee21} over a broad spectral range by repeating similar campaigns at multiple epochs.

\begin{longrotatetable}
	\begin{deluxetable*}{cc|cccc}
\tabletypesize{\scriptsize}
\tablecaption{Summary of the campaign observations \label{tab:obs_summary}}
\tablewidth{0pt}
\tablehead{\colhead{Location} & \colhead{Facility/Inst.} & \colhead{Type*} & \colhead{Spectral range [nm]} & \colhead{Date} & \colhead{Status \& Sect.}}
\decimalcolnumbers
\startdata
{ Space (Venus orbit) } & Akatsuki/UVI & I & 283, 365 & Regular monitoring & Success, Sect.~\ref{subsec:uvi} \\
{ Space (interplanetary) } & BepiColombo/PHEBUS & S & 145--315, 402, 423 & Aug~28--Sep~2 & Insufficient for analysis Sect.~\ref{subsec:phebus} (not used) \\
{ Space (Earth orbit) } & Hisaki/EXCEED & S & 52-148 & Aug~21--Sep~3 & Success for relative analysis (not used) \\
\cline{1-6}
{ Spain } & CAHA1.23/DLR-MKIII & I & BVRI bands & Aug~21--28 & Success, Sect.~\ref{subsec:caha1.23} \\
{ Spain } & CAHA2.2/PlanetCam & I & 380-1700 & Aug~28--31 & Insufficient for analysis (not used) \\
{ Spain } & CAHA3.5/PMAS & I\&S & 364--457 ($d\lambda=$0.28~nm) & Aug 27--30 & Success for relative analysis, Sect.~\ref{subsec:pmas} \\
{ Turkey } & TUG/T100 & I & UBV bands & Aug~25--Sep~2 & Insufficient for analysis (not used) \\
{ Spain (Tenerife) } & STELLA/WiFSIP & I & U band & Aug--Nov & Success, Sect.~\ref{subsec:stella} \\
{ Czech Republic } & Perek telescope/OES & S & 375.3--919.5 & Aug~21--Sep~2 & Insufficient for analysis (not used) \\
\enddata
\tablecomments{*I: Image, S: Spectrum}
\tablecomments{The list of acronyms\\UVI: UltraViolet Imager\\PHEBUS: Probing of Hermean Exosphere By Ultraviolet Spectroscopy\\EXCEED: EXtreme ultraviolet spectrosCope for ExosphEric Dynamics\\CAHA: Calar Alto observatory\\PMAS: Potsdam Multi-Aperture Spectrophotometer\\TUG: TUBITAK National Observatory\\WiFSIP: Wide-Field STELLA Imaging Photometer\\OES: Ondrejov Echelle Spectrograph}
\end{deluxetable*}
\end{longrotatetable}

\section{Data} \label{sec:data}
Details of data acquisition and calibrations are described in this section for each instrument.

\subsection{Akatsuki/UVI} \label{subsec:uvi}
UVI has two filters centered at 283 and 365~nm \citep{Yamazaki18}. The 365~nm wavelength is to detect the absorption by the unknown absorber, and the 283~nm wavelength is located near the center of a SO$_2$ band. In the regular observation mode, UVI obtains Venus images at the two filters every 2 hours from a highly elliptical equatorial orbit. We selected images with the complete coverage of the Venus dayside between 2015 December 7th and 2021 March 31st. Some known artifact images are excluded in the data set.

In this analysis we used two flat-fields; the first flat-field was measured in a laboratory before the launch \citep{Yamazaki18} and a new flat-field that has been prepared with the diffuser images acquired in 2020-2021. The first flat-field was applied to images before 2019 September 17th, and the new flat-field was applied to images from 2019 September 17th. Both flat-fields are publicly available in the calib directory of DARTS data sets (http://darts.isas.jaxa.jp/pub/doi/VCO-00016.html). Using star observations in 2010--2020, the calibration correction factors ($\beta$) were calculated. The averaged $\beta$ are 1.533$\pm$0.208 at 365~nm, and 1.991$\pm$0.279 at 283~nm. These $\beta$ are close to the values reported in \citet{Yamazaki18}. We notice a weak sensitivity change with time at 283~nm, but not evident at 365~nm. Star observations by UVI will continue, so we will examine possible sensitivity changes more in detail in the near future. In this study, we took the averaged $\beta_{\lambda}$ for each channel ($\lambda$).

We calculated the disk-integrated flux of Venus, $F_{{\rm Venus}}$ in [W m$^{-2} \mu$m$^{-1}$], as follows:
\begin{equation}
	\centering
	F_{{\rm Venus}}(\alpha,\lambda,t)=\sum_{r<r_o}\beta_{\lambda}I(x,y)\times\Omega_{\rm pix}
	\label{eq:F_venus}
\end{equation}
where $\alpha$ is the phase angle, $\lambda$ is the wavelength, $t$ is the observation time, $I$ is the measured radiance at $(x,y)$ pixel locations on an image, $\Omega_{\rm pix}$ is the solid angle of one pixel, and $r$ is the distance of ($x,y$) from the Venus disk center. $r_o$ is the limiting distance of integration that includes the Venus radius in pixels and the point spread function (7~pixels). So $r<r_o$ defines an area of flux integration from the planet center ($r$=0) to $r_o$. Then, we subtracted the mean background noise per pixel. The solid angle of Venus, $\Omega_{\rm Venus}$($t$), was calculated as,
\begin{equation}
	\centering
	\Omega_{\rm Venus}(t)=\pi \left(\arcsin\left(\frac{R_{\rm {Venus}}}{d_{\rm V-obs}(t)}\right)\right)^2,
	\label{eq:Omega_venus}
\end{equation}
where  $R_{\rm {Venus}}$ is the radius of Venus and $d_{\rm V-obs}$ is the distance of the spacecraft from Venus in km at the time of observation $t$. For $R_{\rm {Venus}}$, we considered the cloud top altitude from the center of the planet (6052+70~km)

We calculated the disk-integrated albedo $A_{\rm disk-int}$, as the following \citep{Sromovsky01},
%\begin{linenomath*}
	\begin{equation}
		\centering
		A_{\rm disk-int}(\alpha,\lambda,t)=\frac{\pi}{\Omega_{\rm Venus}(t)}\frac{{d_{\rm V-S}(t)}^2 F_{\rm Venus}(\alpha,\lambda,t)}{S_{\odot}(\lambda)},
		\label{eq:APhi}
	\end{equation}
%\end{linenomath*}
where $d_{\rm V-S}(t)$ is the distance between Venus and the Sun [au] at the time of observation $t$, $\Omega_{\rm Venus}$($t$) is the solid angle of Venus as viewed from Akatsuki, and $S_{\odot}(\lambda)$ is the solar irradiance at 1 au [W m$^{-2} \mu$m$^{-1}$] (see, Sect.~\ref{subsec:sun}), calculated for the transmittance functions of each filter. $A_{\rm disk-int}$ is similar in meaning to the radiance factor \citep{Hapke12} that can be applied to spatially resolved images. $A_{\rm disk-int}(\alpha=0^{\circ},$ $\lambda)$ is the `geometric albedo' at wavelength $\lambda$.

\begin{figure*}
	\gridline{
		\fig{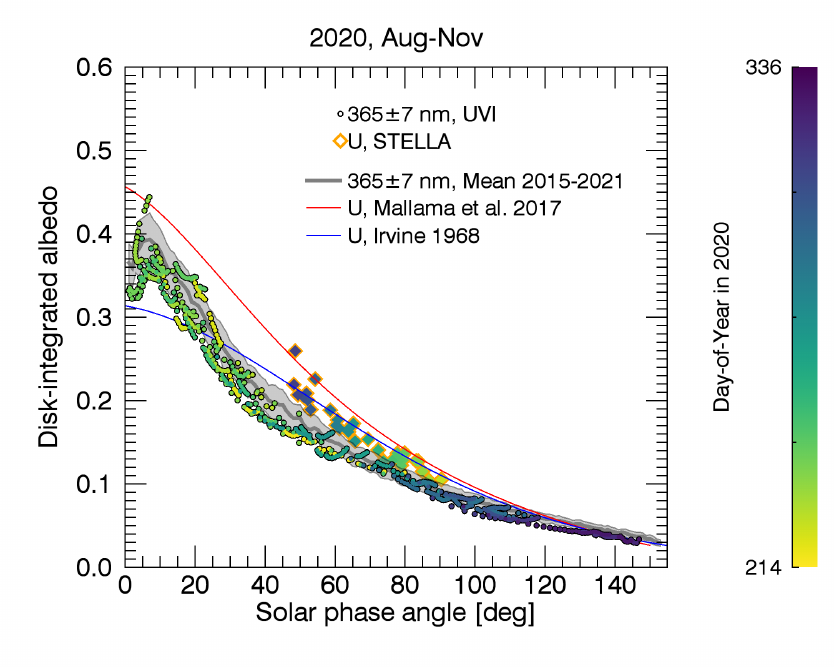}{0.6\textwidth}{(a)}
	}
	\gridline{
		\fig{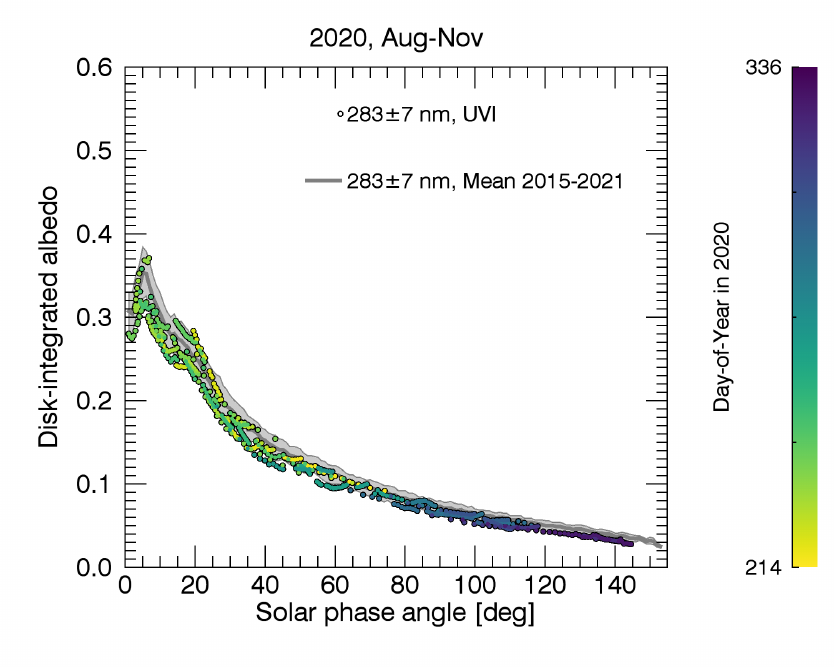}{0.6\textwidth}{(b)}
	}
	\caption{Observed disk-integrated albedo at (a) 365~nm and (b) 283~nm as a function of phase angle. The gray solid lines are the mean phase curve of Akatsuki/UVI, and the light gray filled areas are the standard deviations. Circles are Akatsuki/UVI data and orange diamonds are the ground-based U band data (Sect.~\ref{subsec:stella}). The filled color of symbols indicates observation date between August 1st and November 30th in 2020 as shown in the colorbar. The red curve in (a) is taken from \cite{Mallama17}, converted from magnitude to albedo. The blue curve is taken from \cite{Irvine68}, also converted from magnitude to albedo. Their original magnitudes are shown in Fig.~\ref{fig:U_phasecurve}.\label{fig:UV_phasecurve}}
\end{figure*}

Fig.~\ref{fig:UV_phasecurve} shows the mean phase curves at the two channels in 2015--2021 (gray lines). The colored circle symbols indicate the data in 2020 August-November, when our ground-based U band observations were conducted (see Sect.~\ref{subsec:stella} for details). The symbols show consistent phase angle dependence within the standard deviations of the mean phase curve (light grey area). The ground-based U band has a wider bandwidth (34~nm) than that of UVI (14~nm), which may be the reason for the systematic offset from the UVI data. Previously reported mean phase curves at U band are compared in the same plot. \citet{Irvine68}'s U band has the widest bandwidth (116~nm). \citet{Mallama17} assumed the same phase angle dependence of the B band, and adjusted the geometric albedo to match with previous observations. Details about the U band are described in Sect.~\ref{subsec:stella}.

\subsection{BepiColombo/PHEBUS} \label{subsec:phebus}
BepiColombo was launched in October 2018, and it is on its way to Mercury (arrival in 2025). BepiColombo is composed of two spacecraft: Mercury Planetary Orbiter (MPO) and Mercury Magnetospheric Orbiter (Mio). PHEBUS is the UV spectrometer on board MPO. BepiColombo made two Venus flybys in October 2020 and August 2021, which became opportunities for close-up observations of Venus \citep{Mangano21}. During the Venus flybys, PHEBUS acquired data over the night side and limb, because the dayside of Venus was too bright for the PHEBUS sensor which is designed to detect faint UV emissions by atmospheric gases of Mercury and the night side albedo of Mercury \citep{Quemerais20}.

The observations of the Venus dayside used here were obtained from a long distance when the tiny planetary disk entered the slit of PHEBUS. On 2020 August 28th$-$September 2nd, there were such opportunities: the 66$''$ arcsec apparent size of Venus was within the $2^{\circ}\times0.2^{\circ}$ FOV (Fig.~\ref{fig:campaign_geometry}), and the PHEBUS team made the first Venus faraway observations. 180--181 images were acquired daily over the consecutive 6 days at the far ultraviolet (FUV, 145--315~nm) and two near ultraviolet (NUV, 404 and 422~nm) detectors. The data acquisition was done at 4550V for the Micro-Channel Plate intensifier, which alters the gain \citep{Chassefiere10}. Dark and effective areas are also measured at the 4550V in flight.

While Venus was successfully captured by PHEBUS for 6 consecutive days, we faced three problems. (1) The first problem was unrealistic fluctuations of photon counts that varied day-to-day. These fluctuations were later found to be caused by pointing accuracy. The observations aimed to put the disk at the center of the FOV, but the spacecraft attitude could not put Venus at the center as planned. Instead, Venus was sometimes located near the boundary of the FOV according to the later examination, resulting in significant reduction of the photon counts. This problem prevents the absolute flux analysis, however it should be fine for relative spectral analysis. (2) The second problem was dark count estimation. The dark measurement (deep space imaging) at 4550V was done a month earlier. As the dark count rate changes with the temperature of the detector, this time difference caused insufficient dark subtraction from the Venus images. The PHEBUS team therefore tried to estimate the dark current using the photon counts over the deep space pixels outside of Venus illuminating area. We confirmed day-to-day consistent patterns, although this may have introduced additional small errors. (3) The third problem was the effective area retrieval at 4550V, which were determined with observations of Spica on 2020 February 4th. The retrieved effective area was as expected at wavelengths shorter than 270~nm, but at longer wavelengths it turned out to be insufficient to get reliable results. This third problem became critical, as this means that we cannot compare the brightness with UVI data at 283~nm, and we cannot quantify either the relative absorption by the SO$_2$ gas over the 240$-$315~nm wavelength range (see Sect.~\ref{sec:discussion} for details).

After the examination explained above, we excluded the PHEBUS data from the scientific analysis in this paper. Looking into the future, PHEBUS should provide valuable information to retrieve the disk-mean SO$_2$ gas abundance, and to understand the unknown absorber in the FUV spectral range, which are the main goals of the campaign. Future PHEBUS observations will resolve the three problems that we identified during this campaign.

\subsection{CAHA1.23/DLR-MKIII} \label{subsec:caha1.23}
The DLR-MKIII CCD camera installed at the CAHA 1.23m telescope performed Venus observations in the Johnson-Cousin's BVRI bands. From August 22nd to 28th UTC, Venus was visible right before the sunrise. Venus' apparent diameter changed from 22$''$ to 20$''$ during the period. HR2208 was selected as a solar-like reference star; its spectral type is G2V \citep{Stepien96}--G5V \citep{Gray03}, and was sufficiently bright near Venus at the same airmass range as that of Venus. The photometric variability of the star is reported to be 0.03 and 0.035 magnitude at V and B, respectively, with a 7.8-day period \citep{Stepien96}. This level of variation has a negligible impact on this study as our accuracy does not reach such level; this is comparable to the daily standard deviations of our measurements.

Venus images were taken under strongly defocused conditions \citep{Gillon09,Southworth09} to spread photons of Venus over the wide FOV of the CCD camera. This successfully prevented saturation of Venus images without a neutral density filter. This benefits accurate flux measurements of Venus. Star observations were done at the normal focus position. One observation cycle was composed of Venus and the star imaging at the four filters (at least 4 images per filter per object), and this cycle was repeated 3--4 times each night.

Usual aperture photometry was used to determine the aperture sizes to integrate fluxes of Venus and the star, and we calculated the Signal-to-Noise ratio (SNR) at the corresponding aperture sizes with the CCD equation. The typical SNR of Venus is $\sim$10$^5$ and that of the star is 1000--2000. Atmospheric extinction coefficients were determined at each filter using a linear regression between the instrumental magnitude of the star and airmass. The range of airmass were 1.6--2.1 each night for both Venus and the star. The atmospheric extinction coefficients were consistent for the first 5 nights. During the last 2 nights partial clouds entered the view, resulting in temporally variable telluric opacity. Since the instrumental magnitude at zero airmass is known to be stable for the CAHA1.23 DLR-MKIII camera, we could compute the instantaneous extinction coefficient at the time of the Venus observations by interpolation also during non-photometric nights, thanks to the repeated cycles between Venus and the star. The apparent magnitude of Venus was calculated with the interpolated atmospheric extinction coefficient and the known magnitudes of the star (Table~\ref{tab:refstar_image}).

\begin{figure}
\centering
	\gridline{
	\fig{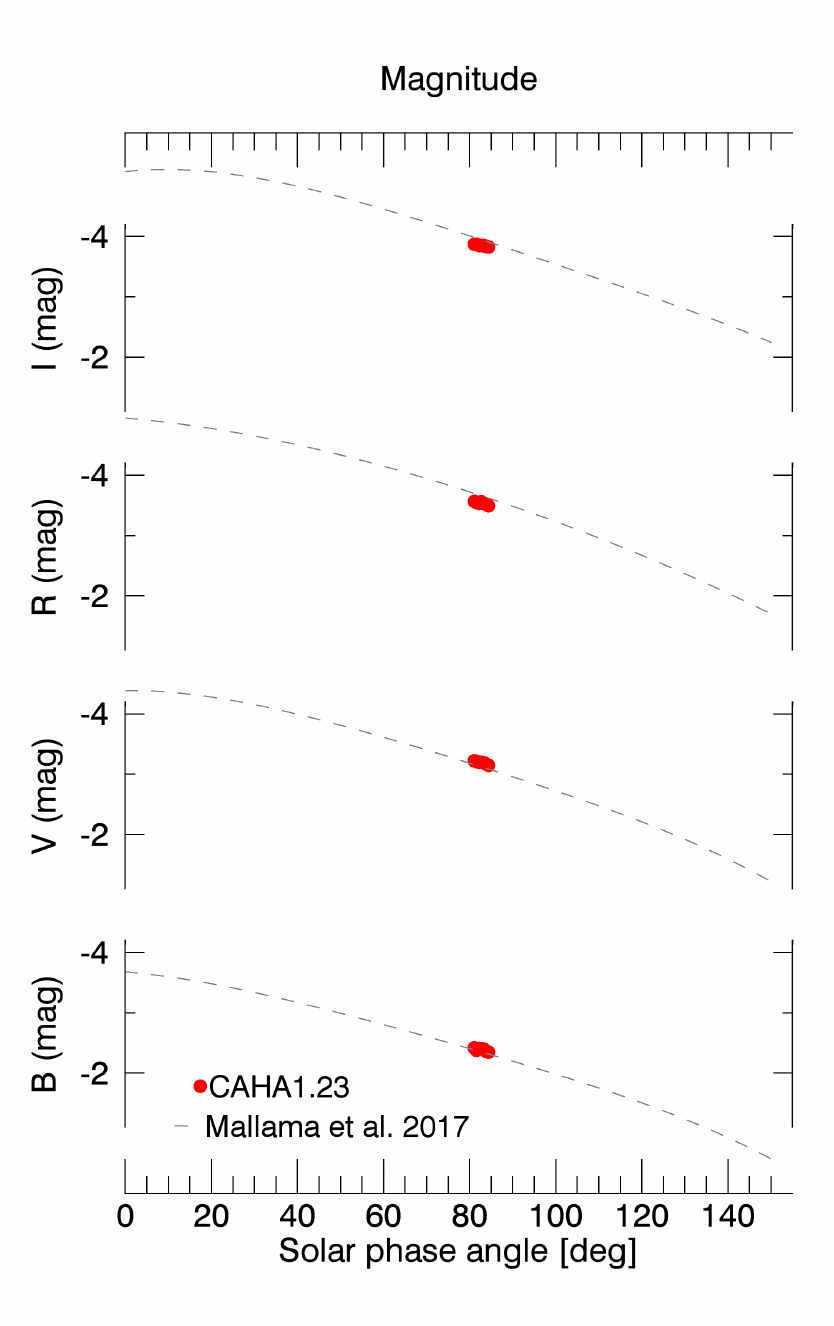}{0.4\textwidth}{(a)}
	\fig{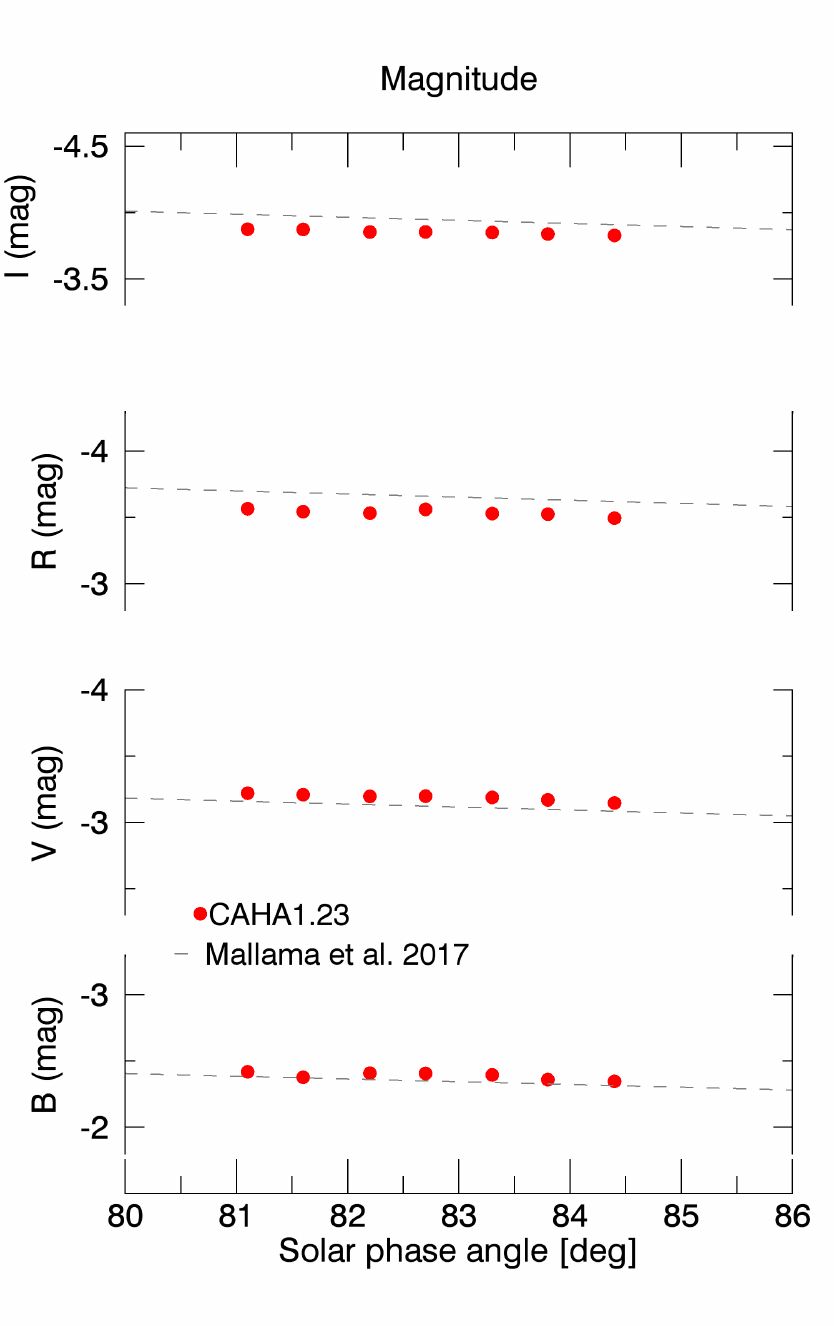}{0.4\textwidth}{(b)}
}
\caption{Observed magnitudes of Venus at Johnson-Cousin's BVRI bands on 2020 Aug 22--28 by the CAHA1.23/DLR-MKIII camera (red dots). (a) Phase curves over the range from 0 to 150$^{\circ}$. (b) Same as (a), but close-up of the observed data points of the campaign. Phase curves at each band are shown in gray dashed lines \citep{Mallama17}. \label{fig:BVRI_phasecurve}}
\end{figure}

The apparent magnitude of Venus was converted to the reduced magnitude, which is the brightness at 1~au from both the Sun and the Earth. The distances between the Sun and Venus, and the Earth and Venus at the time of imaging were calculated using the JPL SPICE toolkit \citep{Acton96}. Hereafter magnitude refers to the reduced magnitude and the results are shown in Fig.~\ref{fig:BVRI_phasecurve}. The comparison with the brightness reported by \citet{Mallama17} shows a good agreement of the expected brightness at $\alpha\sim80^{\circ}$ at the four bands (for daily variation, see Fig.~\ref{fig:R_timeseries}).

\begin{deluxetable*}{ccccccccc}
	\tablecaption{Reference stars for imaging observation \label{tab:refstar_image}}
	\tablewidth{0pt}
	\tablehead{\\
		\colhead{Star} & \colhead{Spectral Type} & \multicolumn5c{Magnitude} & \colhead{Reference of magnitude} & \colhead{Dates of observation} \\
		\cline{3-7}
		& & \colhead{U} & \colhead{B} & \colhead{V} & \colhead{R} & \colhead{I} &	
	}
	\decimalcolnumbers
	\startdata
	HR2208 & G2V--G5V & 7.317 & 7.131 & 6.456 & 6.087 & 5.740 & \cite{Stepien96} & Aug~11$-$Sep~4 \\
	$\kappa$ Gemini & G8III-IIIb & 5.19 & 4.49 & 3.57 & 2.86 & 2.41 & \cite{Ducati02} & Sep~6$-$13 \\
	mu.02 Cnc & G1IVb & 6.14 & 5.93 & 5.30 & & & \cite{Ducati02} & Sep~14$-$30 \\
	35 Leo & G1.5IV-V & 6.85 & 6.64 & & & & \cite{Ducati02} & Oct~2$-$10 \\
	HD88725 & G3/5V & 8.34 &  8.33 & 7.73 & 7.24 & 6.89 & \cite{Ducati02} & Oct~14$-$24 \\
	HD92719 & G1.5V & 7.519 & 7.406 & 6.767 & 6.42 & 6.083 & \cite{Koen10} & Oct~27$-$Nov~8 \\
	\enddata
\end{deluxetable*}

\subsection{STELLA/WiFSIP} \label{subsec:stella}
The WiFSIP wide-field imager installed at the STELLA 1.2m robotic telescope conducted Venus imaging at the U band. The period of observations continued between 2020 August 11th and November 8th, except the time when a Sahara dust storm affected the telescope site in Tenerife. The observations were done before sunrise every day. For about 30~min a bright solar-like reference star near Venus was continuously observed to define the telluric extinction coefficient. Then immediately Venus imaging followed when Venus rose high in the dark sky. Typically 15 Venus images were acquired each night (except Aug 12th, when 5 images were acquired). The airmass of Venus changed with time of observations: 1.6$-$1.7 on August 11--September 14, 1.7$-$1.8 on September 20--October 2, 1.8$-$2.0 on October 3--27, and 2.0$-$2.4 until November 8.

Aperture photometry was applied to determine the size of area to integrate the Venus flux and reference stars. The typical SNR of Venus is 6000--8000, and those of stars are from $\sim$500 to $\sim$3000, depending on the stars. Following the locations of Venus on the sky, our reference stars were changed with time (Table~\ref{tab:refstar_image}). Note that $\kappa$ Gemini has an accompanying star, and its corresponding pixels were excluded in the aperture photometry. The ranges of star airmass varied with time, e.g., 1.9--2.2 on August 11st, 1.4--1.9 on September 14th, 1.5--2.0 on October 10th, and 1.6--1.75 on November 8th. Daily extinction coefficients were monitored and we excluded dates of abnormal behavior compared to other dates. Our averaged extinction coefficient at the U band is 0.485$\pm$0.093. The apparent magnitude of Venus was calculated with the daily atmospheric extinction coefficient and the known magnitudes of each star (Table~\ref{tab:refstar_image}). The apparent magnitude was converted to the reduced magnitude as described in Sect.~\ref{subsec:caha1.23}.

\begin{figure}
	\plotone{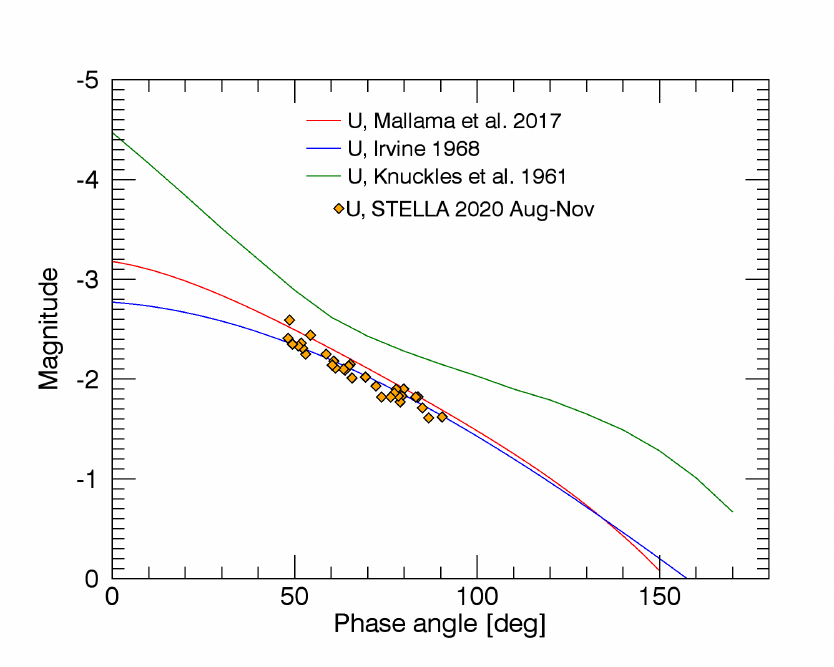}
	\caption{Observed magnitude of Venus at Johnson U band by STELLA/WiFSIP in 2020 August$-$November (orange diamonds). Phase curves reported in previous studies are compared: \cite{Knuckles61} (green), \cite{Irvine68} (blue), and \cite{Mallama17} (red). \label{fig:U_phasecurve}}
\end{figure}

Our STELLA U band magnitude measurements are to the best of our knowledge the first after \citet{Irvine68}. The comparison of these data sets is shown in Fig.~\ref{fig:U_phasecurve}. As a reference, two more data sets are shown together: the oldest measurement \citep{Knuckles61} and a recent estimation \citep{Mallama17}. The comparison of our data with \citet{Irvine68} shows a consistent magnitude, but it is in fact an inadequate comparison considering the wider bandwidth of \citet{Irvine68} (116~nm) than that of STELLA (34~nm). \citet{Mallama17} estimated the U band phase curve that follows the phase angle dependency at the B band and has the geometric albedo to be consistent with the two older U band observations. \citet{Knuckles61} shows a much brighter Venus magnitude, and it is difficult to understand what causes such a difference. In this study, we adopted the \cite{Irvine68}'s phase curve as a reference phase curve at U to correct the phase angle dependence of STELLA data (Eq.~\ref{eq:rel_A} in Sect.~\ref{sec:results}).

The fluctuation of STELLA's U band is noticeable in Fig.~\ref{fig:U_phasecurve}. This may be real short-term fluctuations as reported in a recent study of the Venus' disk-integrated albedo \citep{Lee20} (see Sect.~\ref{subsec:result_obs_timeseries}). Venus monitoring will continue by STELLA, and we should be able to construct the true mean phase curve at U and extract accurate temporal variations in the near future.

\subsection{CAHA3.5/PMAS} \label{subsec:pmas}
\begin{figure*}
	\gridline{
		\fig{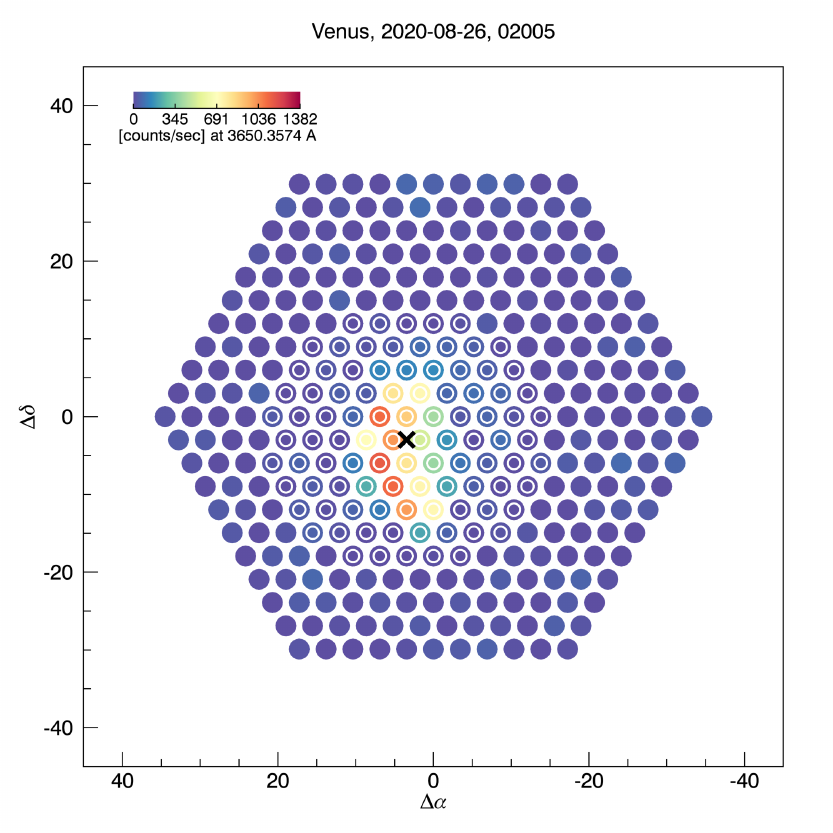}{0.4\textwidth}{(a)}
		\fig{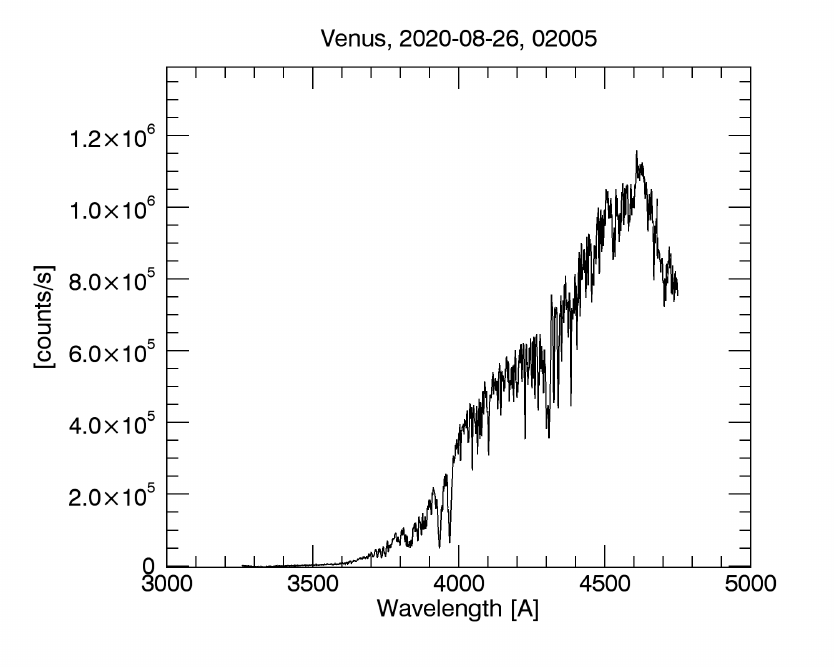}{0.4\textwidth}{(b)}
	}
	\gridline{
		\fig{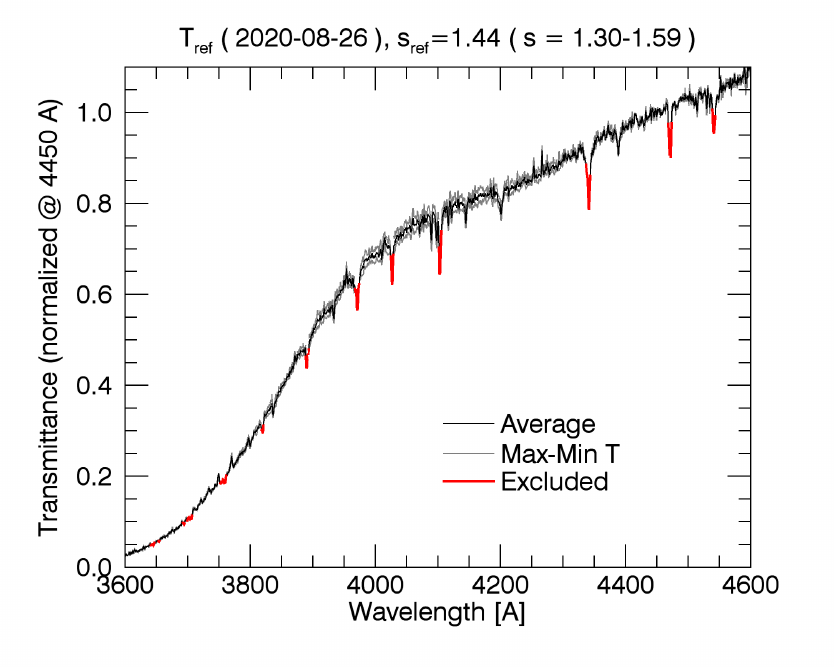}{0.4\textwidth}{(c)}
		\fig{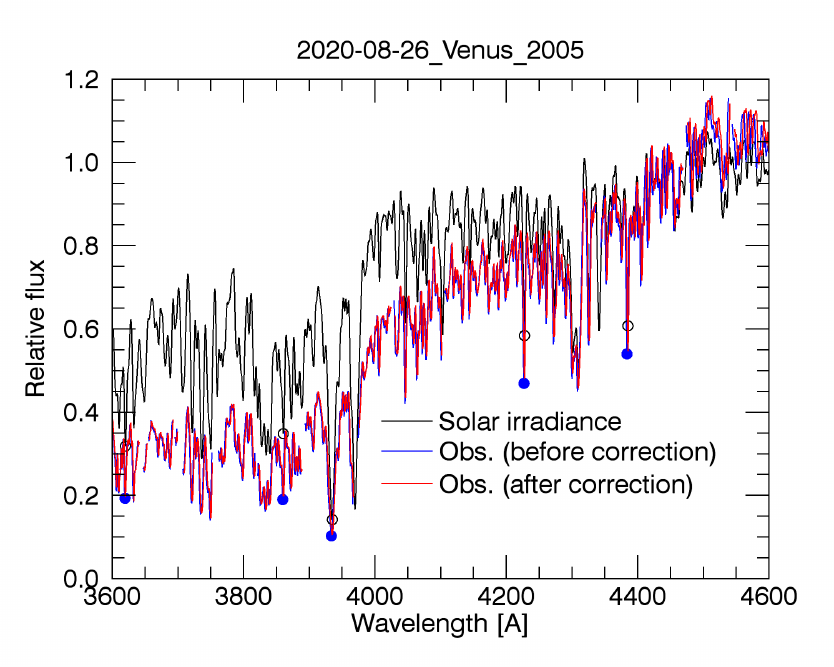}{0.4\textwidth}{(d)}
	}
	\caption{PMAS observation and calibration examples of the data acquired on 2020 Auguest 26th (image number 2005). (a) Image slice of PPAK at 365~nm after the sky subtraction. Strong signals are where Venus dayside is located. The mean of strong signal location is marked with the black `X', from which the 16$''$ radius circular area is selected for the flux integration (`$\circ$' symbols). (b) Integrated flux spectrum of Venus before the telluric extinction correction. (c) Relative telluric transmittance function $T_{\rm{ref}}(\lambda)$ of the same night (see text for the details). The star (10Lac) was observed at the airmass $s$ of 1.30 and 1.59 (grey curves). The mean transmittance is used to define $T_{\rm{ref}}(\lambda)$ (black), except fine emissions of the star (red ranges). (d) Relative flux spectrum of Venus $F_{\rm{Venus}}(\lambda)$ after the telluric extinction correction (blue). Relative solar irradiance is shown (black). Some features are selected (black circles) as spectral references to adjust the spectral locations of Venus flux (blue disks). After this fine spectral location correction, the final Venus spectrum is shown in the red curve, which is almost overlapped with the blue curve. \label{fig:PMAS_data}}
\end{figure*}

The Potsdam Multi-Aperture Spectrophotometer (PMAS) is installed at the CAHA 3.5m telescope \citep{Roth05}, and acquired Venus data on 2020 August 26--29. Venus observations were done with the bare fiber bundle integral field unit (PPAK) which has a wide hexagonal FOV of $65''\times74''$ as shown in Fig.~\ref{fig:PMAS_data}a. A total of 331 fibers obtained scientific data within the FOV, and additional 36 fibers simultaneously acquire sky data at 72$''$ away from the center of the FOV. The wide FOV is sufficient to capture the entire Venus disk that had the apparent diameter of $\sim20''$.

PMAS is optimized to observe faint objects. In fact Venus is too bright for PMAS in the normal operation mode, so the Venus observation was conducted with special care; only one petal of the mirror cover was open to reduce the photon flux. Our target spectral range of the observation, from UV to blue color, also helped to reduce the photon flux of Venus due to the telluric extinction. We acquired high spectral resolution data, $d\lambda=$0.28~nm, which effectively spread the photons between 326.1 and 478.3~nm using the U1200 grating.

Venus data were acquired at the end of each night for about an hour with the 0.4~s exposure time, resulting in $\sim$100 images per night. The airmass of Venus ranged typically between 1.5 and 2.5 each night, and only data for which the airmass was close to that of our reference star observations were selected for the analysis. Reference star observations were done at the beginning of night, in the middle, and right before the Venus observations. 10Lac, Vega, and eps Aqr were observed each night, and we eventually used 10Lac (CALSPEC database\footnote{\url{https://archive.stsci.edu/hlsps/reference-atlases/cdbs/current_calspec/10lac_mod_003.fits}}, \citet{Bohlin14}) to construct the telluric transmittance function that makes use of PMAS' high spectral resolution. Two sets of 10Lac observations were done each night between 1.3 and 1.7 airmass.

The P3D version 2.7 package was used for the data reduction \citep{Sandin10}, which includes creating a master bias image, tracing spectra, finding spectral positions, and generating a flat-field; and produces a reduced spectral image. We integrated the flux of the targets as follows \citep{Rosales-Ortega10}. We subtracted the median sky spectrum from the scientific data in each image. From the center of Venus or the reference star (e.g., the black `X' mark in Fig.~\ref{fig:PMAS_data}a), we increased the area of flux summation until the total flux does not increase further ($<$1\%) over the entire wavelength range. Such radius was 16$''$ for Venus and 12$''$ for 10Lac. The selected fibers of the flux integration area is marked with white circles in Fig.~\ref{fig:PMAS_data}a, and the total flux spectrum of Venus is shown in Fig.~\ref{fig:PMAS_data}b. 

We generated a reference telluric transmittance function $T_{\rm{ref}}(\lambda)$ each night \citep{Wyttenbach15} as
\begin{equation}
T_{\rm{ref}}(\lambda)=\exp(E_{\lambda}s_{\rm{ref}}),
\end{equation}
where $\lambda$ is the wavelength, $E_{\lambda}$ is the telluric optical depth at zenith, and $s_{\rm{ref}}$ is the mean airmass of the reference star observations. The difference from \citet{Wyttenbach15} is that the transmittance is not at the unity airmass (zenith), but at $s_{\rm{ref}}$.

We calculated $E_{\lambda}s_{\rm{ref}}$ as follows. The logarithm of the measured star fluxes $F_{\rm{obs,star}}(\lambda)$ has a linear relationship with the airmass $s$ \citep{Langeveld21} as
\begin{equation}
	\ln(F_{\rm{obs,star}}(\lambda))=E_{\lambda}s+c,
\end{equation}
where c is a constant. As the standard star flux spectrum $F_{\rm{std}}(\lambda)$ is at $s=0$, we can convert the equation above as
\begin{equation}	 
	\ln({F_{\rm{obs,star}}(\lambda)})-\ln({F_{\rm{std}}(\lambda)})=E_{\lambda}s+c_0,
\end{equation}
where $c_0$ is a constant to make the result to be zero at 445~nm, where is the center of the Johnson B band. We normalized transmittance functions to that at 445~nm, and averaged two sets of normalized transmittance functions. The $T_{\rm{ref}}(\lambda)$ on August 26th is shown in Fig.~\ref{fig:PMAS_data}c (black curve). Fine emission lines of the star are excluded in this process (red intervals).

We retrieved relative flux spectra of Venus $F_{\rm{Venus}}(\lambda)$ using $T_{\rm{ref}}(\lambda)$; after selecting observed Venus fluxes $F_{\rm{obs,venus}}(\lambda)$ that were acquired at $s$ near the star observations ($ds<$0.1) each night, we divided these Venus fluxes by $T_{\rm{ref}}(\lambda)$ of the same night,
\begin{equation}
	F_{\rm{Venus}}(\lambda)=\frac{F_{\rm{obs,venus}}(\lambda)}{T_{\rm{ref}}(\lambda)}.
\end{equation}

An example of relative Venus flux spectrum on August 26th is shown in Fig.~\ref{fig:PMAS_data}d (blue curve, which almost overlaps with the red curve). The comparison between the Venus flux and the solar irradiance (Sect.~\ref{subsec:sun}) is shown in the same figure. Using the selected spectral features of the solar reference (circle symbols), we adjusted slightly the spectral location of the Venus spectrum as shown in the same plot, before (blue) and after (red) the adjustment. Such spectral location adjustment were done between $-4$ and $+3$~$\AA$, depending on date of observations and wavelengths. This last process shows only little changes, but helps to remove unrealistic humps in the reflectivity spectrum. We repeated the same procedure for data of each night to generate daily mean Venus spectra.

\subsection{Solar irradiance data} \label{subsec:sun}
We converted the observation data to reflectivity using the reference solar irradiance spectrum. We used the observed solar irradiance data by TSIS-1 SIM (Version 6, Level 3, Daily data\footnote{\url{https://lasp.colorado.edu/home/tsis/data/}}) over the 200--2400~nm wavelength range. For the campaign data of this paper, we averaged the TSIS-1 SIM data from August to September of 2020. This mean solar spectrum was used to calculate the solar irradiance at 283 and 365~nm for the Akatsuki data. We calculated the solar magnitude at the U band using the effective transmittance function of STELLA, following the description in \citet{Willmer18} to take into account its narrow band width (FWHM=34~nm). For the BVRI broad bands, we took the values given in \citet{Willmer18}. The solar magnitudes at each band are listed in Table~\ref{tab:solmag}.
For the high resolution spectral grids of the PMAS data, that of the TSIS-1 SIM data was not sufficient (5~nm at $\lambda$$\sim$400~nm)\footnote{\url{https://lasp.colorado.edu/home/tsis/instruments/sim-spectral-irradiance-monitor/}}, and we took the SAO2010 solar reference spectrum, whose spectral resolution is 0.04~nm (FWHM) \citep{Chance10}. We convolved the SAO2010 spectrum into the spectral grids of the PMAS data with a Gaussian function, which is shown in Fig.~\ref{fig:PMAS_data}d.

\begin{deluxetable*}{ccccc}
	\tablecaption{Solar magnitude \label{tab:solmag}}
	\tablewidth{0pt}
	\tablehead{\\
		\colhead{U} & \colhead{B} & \colhead{V} & \colhead{R} & \colhead{I}
	}
	\startdata
	-26.0 & -26.13 & -26.76 & -27.15 & -27.47 \\
	\enddata
\end{deluxetable*}

\section{Atmospheric structure and model calculations} \label{sec:model}
We performed radiative transfer model calculations to compare with the observed reflectivity of Venus. The atmospheric structure and the model configurations are described in this section. 

\subsection{Atmospheric gases} \label{subsec:atm}
We took into account atmospheric gaseous absorption of CO$_2$, SO$_2$, OCS, O$_3$, SO, H$_2$O, H$_2$S, HCl, HF, and CO. We also considered CH$_4$, which was tentatively detected by the Pioneer Venus Large probe Neutral Mass Spectrometer \citep{Donahue93}, although the detection is questionable as stated by the authors. Subsequently, CH$_4$ was excluded from our analysis (Sect.~\ref{sec:results}) due to its too strong absorption signature in NIR (Appendix~\ref{Appendix}), which cannot be missed in spectral observations, for example, the spectrum of MESSENGER/MASCS as shown in \citet{Perezhoyos18}. The similar conclusion is also drawn from the night side spectral data analysis at the 2.3~$\mu$m atmospheric window, providing the upper limit of methane as $<$0.1~ppm at 30~km altitude \citep{Pollack93}. The detection of CH$_4$ may require future observations to confirm its possible presence \citep{Johnson19,Bainse21}, and this is beyond the concern of this study.

Altitudes between 48 and 100~km were modeled in this study, including the sulfuric acid cloud layer (Sect.~\ref{subsec:clouds}). Molecular number density was calculated as a function of altitude using the atmospheric temperature and pressure profiles at low latitudes \citep{Seiff85}. Gaseous absorption has been calculated as a function of either temperature (especially for the UV absorption cross-sections), or altitude that considers both temperature and pressure of Venus (for the line-by-line calculations).

We compared the available cross-section data sets in the UV-visible wavelength range of SO$_2$, OCS, O$_3$, CO$_2$, SO, H$_2$O, H$_2$S, and HCl, and finally selected those shown in Table~\ref{tab:gas_data}. We selected data published in recent years and measured experimentally, or recommended by another data base, such as the JPL compilation by \citet{Burkholder20}. SO$_2$, OCS, and CO$_2$ absorption is known to depend on temperature. We assumed linear functions of temperature for the cross-sections of SO$_2$ \citep{Vandaele09} and OCS \citep{Grosch15}. That of CO$_2$ was calculated with exponential functions, following \citet{Hartinger00} and \citet{Venot18}.

\begin{figure*}
	\gridline{
	\fig{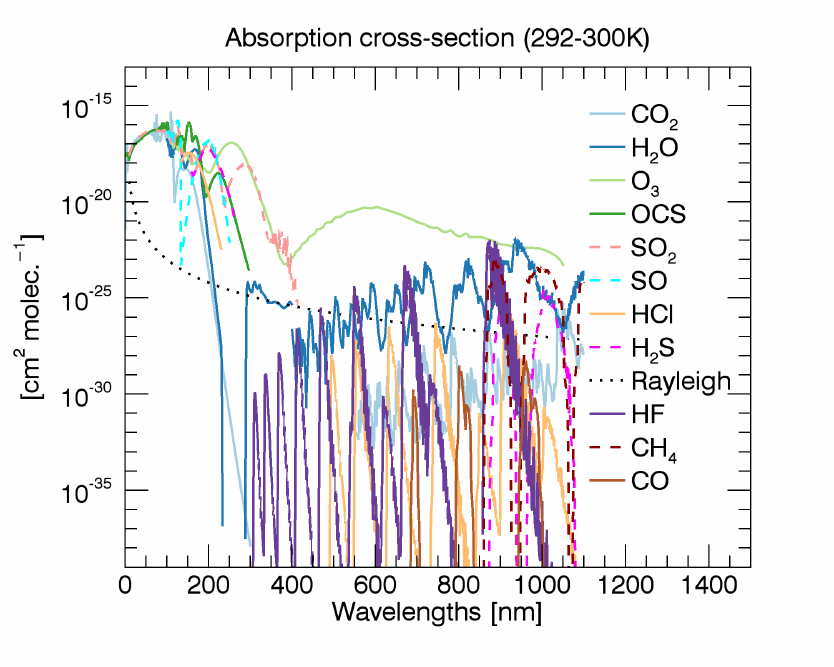}{0.5\textwidth}{(a)}
	\fig{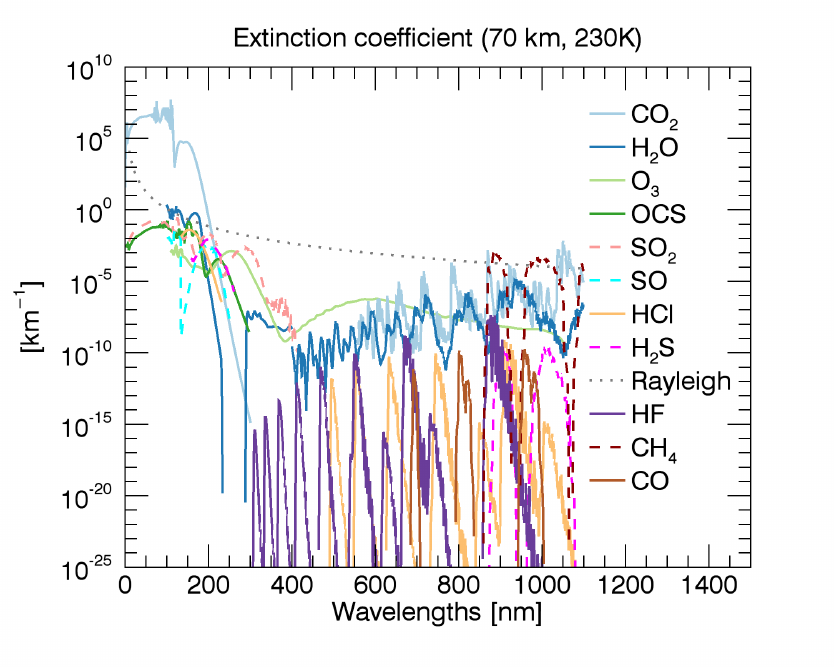}{0.5\textwidth}{(b)}
    }
	\gridline{
	\fig{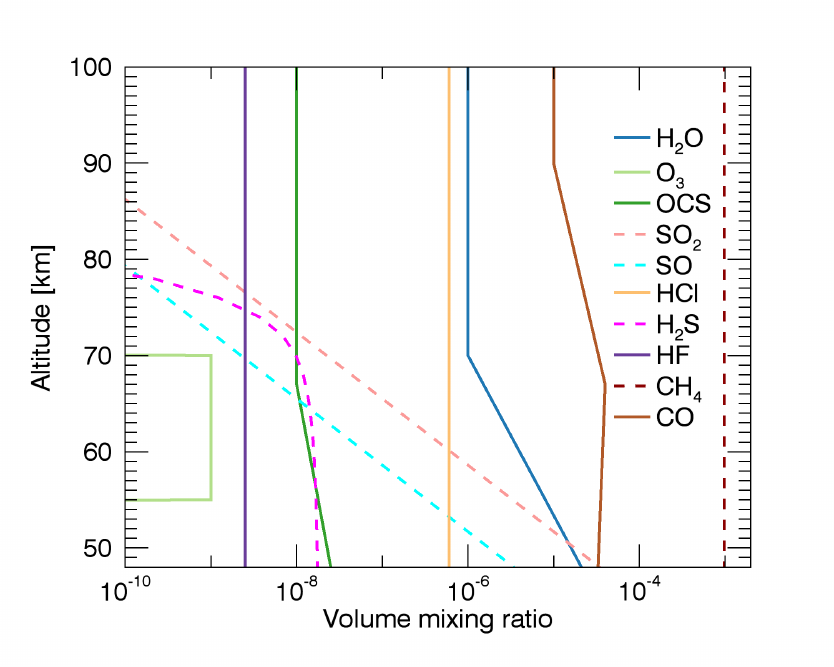}{0.5\textwidth}{(c)}
	}
	\caption{Atmospheric gases. (a) Absorption cross-section of gases considered in this study at room temperature. Rayleigh scattering is also plotted for comparison (dotted line). Dashed lines indicate gases that require careful consideration due to its highly variable abundance (SO$_2$ and SO) or its limited observations (H$_2$S and CH$_4$). (b) Extinction coefficient of gases at the 70~km altitude atmosphere (near the cloud top level). (c) Vertical profiles of gaseous abundances assumed in this study. \label{fig:atm_gases}}
\end{figure*}

At the longer wavelengths, from visible to NIR wavelength, we calculated line-by-line cross-sections with a 0.1~cm$^{-1}$ spectral resolution: \citet{Huang17} for CO$_2$, and HITRAN2016 \citep{Gordon17} for H$_2$O, H$_2$S, HCl, HF, and CO. The sub-Lorentzian factor of the CO$_2$ absorption was assumed to be that of the CO$_2$ atmospheric windows between 1.18 and 2.3~$\mu$m, produced a reasonable fit with observations \citep{Meadows96}. We used the 300~cm$^{-1}$ line-cutoff value, which is longer than 200~cm$^{-1}$ in a previous study  \citep{Lee16}, instead of assuming possible CO$_2$ continuum. All other gases used the 100~cm$^{-1}$ line-cutoff value without a sub-Lorentzian factor assumption. Rayleigh scattering was calculated for the atmosphere composed of 96.5\% CO$_2$ and 3.5\% N$_2$ with cross-sections by \citet{Sneep05}. A summary of the gaseous absorption cross-section data at the room temperature is shown in Fig.~\ref{fig:atm_gases}a. The cross-section of the Rayleigh scattering is also plotted for comparison. Figure~\ref{fig:atm_gases}b shows extinction coefficients of gases at the cloud top atmosphere ($\sim$70~km) with the assumed abundances of the trace gases (Fig.~\ref{fig:atm_gases}c). In Fig.~\ref{fig:atm_gases}b the dominant CO$_2$ absorption is noticeable at the short wavelength edge, while the Rayleigh scattering dominates over the other wavelengths. We note that SO$_2$ and O$_3$ absorption bands overlap around 250~nm. We also note that SO, SO$_2$, and H$_2$S absorption overlap at $\lambda<$240~nm. H$_2$S, if its presence is confirmed, may cause a complication to retrieve abundance of the SO gas in low spectral resolution data, and this has not been considered in a previous study \citep{Belyaev12}.

The vertical profiles of gaseous abundance (Fig.~\ref{fig:atm_gases}c) were assumed as follows. CO$_2$ was fixed to 96.5\% \citep{Seiff85}. A recent observation analysis revealed the cloud top level O$_3$ \citep{Marcq19_O3}, and its global mean value was assumed as $\sim$1~ppbv at 55$-$70~km altitude. Chemistry model calculations suggested the possible presence of H$_2$S \citep{Bierson20}, and their nominal model profile was adopted in this study: $\sim$10~ppbv near 70~km altitude and decreasing above. This abundance of H$_2$S is lower than that used in a previous study \citep{Titov07} whose atmospheric model was based on in-situ measurements at lower altitudes ($\le$55~km) \citep{vonZahn85}. Observed SO$_2$ and SO abundances at the cloud top level are highly variable \citep{Marcq19,Encrenaz19}; we assumed SO$_2$ abundance at 70~km as 22.4~ppbv \citep{Lee21} and SO as 10\% of SO$_2$, following \citet{Marcq19}. The vertical distribution of SO$_2$ gas abundance was assumed to change with a scale height of 3~km \citep{Marcq19}. We also assumed HCl as 0.6~ppmv \citep{Connes67}, and followed the assumptions in \citet{Titov07} for OCS, H$_2$O, HF, and CO. While testing the impact of CH$_4$ on the simulated spectrum, 980~ppmv was assumed \citep{Donahue93} (Appendix~\ref{Appendix}).

\begin{deluxetable*}{cccccc}
	\tablecaption{Gaseous absorption dataset \label{tab:gas_data}}
	\tablehead{\\
		\colhead{Gas} & \colhead{Wavelength} & \colhead{Measured temp.} & \multicolumn{2}{c}{Dependence} & \colhead{Reference (data source, if applicable)}\\
            \cline{4-5}
			 & \colhead{[nm]} & \colhead{[K]} & \colhead{Temp.} & \colhead{Pres.} &		
			}
	\decimalcolnumbers
	\startdata
	SO$_2$ & 10--106 & 298 & $\times$ & $\times$  & \citet{Feng99} (MPI-Mainz Atlas$^1$) \\
	       & 106--230  & 293 & $\times$ & $\times$  & \citet{Manatt93} (MPI-Mainz Atlas) \\
	       & 230--417  & 298--358 & $\circ$ (linear) & $\times$  & \citet{Vandaele09,Hermans09} \\
	\hline
	OCS    & 3.44--115 & 298 & $\times$  & $\times$  & \citet{Feng00a,Feng00b} \citep{Heays17} \\
	       & 115--190 & 298 & $\times$  & $\times$  & \citet{Limao-Vieira15} \citep{Heays17}  \\
	       & 190--205 & 295 & $\times$ & $\times$ & \citet{Molina81} (JPL recommendation 2000$^2$, MPI-Mainz Atlas) \\
    	   & $\le$315  & 294.8--773.2 & $\circ$ (linear) & $\times$ & \citet{Grosch15} (MPI-Mainz Atlas) \\
	\hline
	O$_3$  & 110--195 & 298 & $\times$  & $\times$  & \citet{Mason96} \\
		   & 197--825  & 293-298 & $\times$  & $\times$  & (JPL recommendation 2000) \\
		   & 853--1047  & 294 & $\times$ & $\times$  & Table~8 in \citet{Helou05} \\
	\hline
	CO$_2$ & 0.1254--106.15 & 300 & $\times$ & $\times$  & \citet{Huestis11} \\
		   & 106.15--115 & 195, 295 & $\circ$$^3$ & $\times$  & \citet{Stark07} \\
	       & 115--230 & 150--800 & $\circ$ & $\times$  & \citet{Venot18} \\
	       & 556--1000 & - & $\circ$ & $\circ$ & \citet{Huang17} \\
	\hline
	SO     & 190--220 & 293 & $\times$ & $\times$ & \citet{Phillips81} \\
	       & 190, 220 & - & $\times$ & $\times$ & Gaussian fit of \citet{Phillips81}$^{4}$ \\
	\hline
	H$_2$O & 99.9--114 & 298 & $\times$ & $\times$ & \citet{Fillion04} (MPI-Mainz Atlas) \\
	       & 114--140 & 298 & $\times$ & $\times$ & \citet{Mota05} (MPI-Mainz Atlas) \\
	       & 140--190 & 298 & $\times$ & $\times$ & (JPL recommendation 2000)	 \\
	       & 192--230 & 292 & $\times$ & $\times$ & The extrapolated model of \citet{Ranjan20} (MPI-Mainz Atlas) \\
	       & 290--325 & 295 & $\times$ & $\times$ & Table~1 in \citet{Pei19} \\
	       & 325--400 & 293 & $\times$ & $\times$ & The upper limits of \citet{Wilson16} (MPI-Mainz Atlas) \\
	       & 400--1000 & - & $\circ$ & $\circ$ & (HITRAN2016$^{5}$) \\
	\hline
	H$_2$S & 160--260 & 170--370 & $\circ$ & $\times$ & \citet{Wu98} (MPI-Mainz Atlas) \\
	       & 833--1000 & - & $\circ$ & $\circ$ & (HITRAN2016) \\
	\hline
	HCl    & 135--230 & 298 & $\times$ & $\times$ & \citet{Bahou01} \\
           & 476-1000 & - & $\circ$ & $\circ$ & (HITRAN2016) \\
	\hline
	HF	   & 303--1000 & - & $\circ$ & $\circ$ & (HITRAN2016) \\
	\hline
	CO     & 667--1000 & - & $\circ$ & $\circ$ & (HITRAN2016) \\
	\hline
	CH$_4$ & 833--1000 & - & $\circ$ & $\circ$	& (HITRAN2016) \\\
	\enddata
	\tablecomments{
	$^{1}$ \citet{Keller-Rudek13}\\
	$^{2}$ \citet{Burkholder20}\\
	$^{3}$ \citet{Hartinger00}\\
	$^{4}$ Gaussian fit of the SO absorption cross section, $\sigma(\lambda)$ [cm$^2$] $= A_0 \exp( -\frac{(\lambda - A_1)^{2}}{2 \times (A_2)^{2}})$, where $\lambda$ is the wavelength in [nm], $A_0=1.15633\times10^{-17}$ cm$^2$, $A_1=194.665$ nm, and $A_2=11.2838$ nm.\\
	$^{5}$ \citet{HITRAN2016}
	}
\end{deluxetable*}

\subsection{Clouds} \label{subsec:clouds}
\begin{figure*}
	\gridline{
		\fig{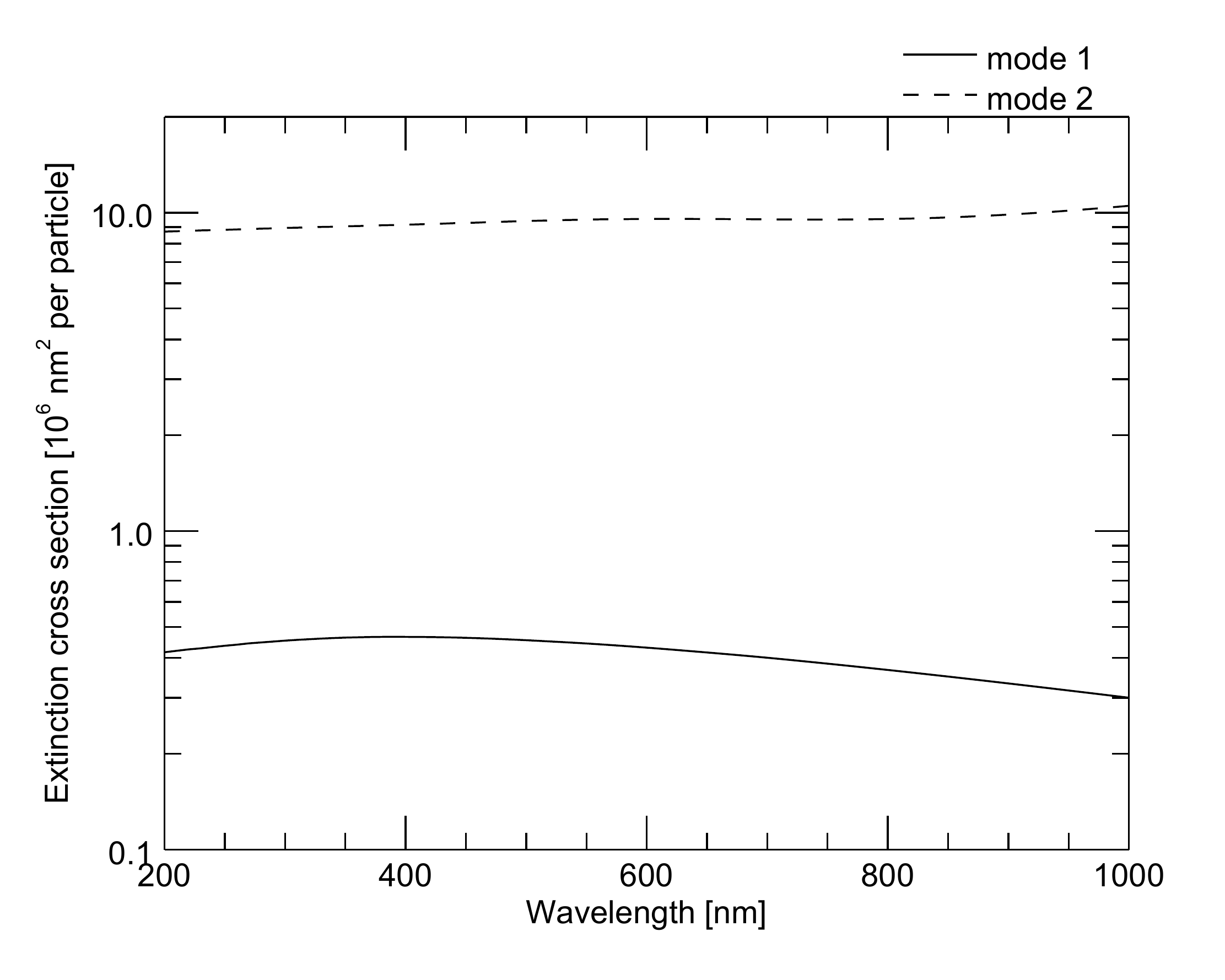}{0.5\textwidth}{(a)}
		\fig{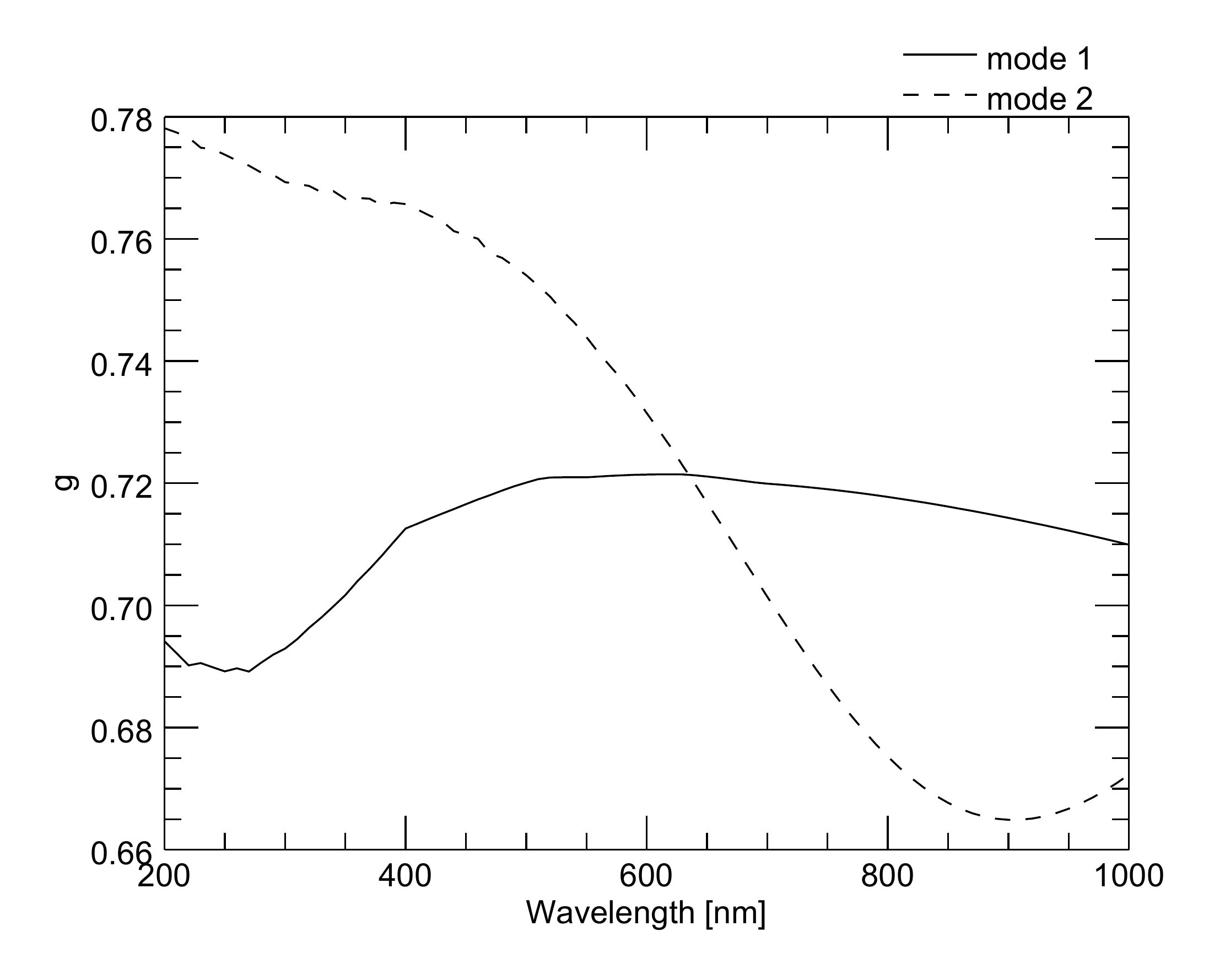}{0.5\textwidth}{(b)}
	}
	\gridline{
		\fig{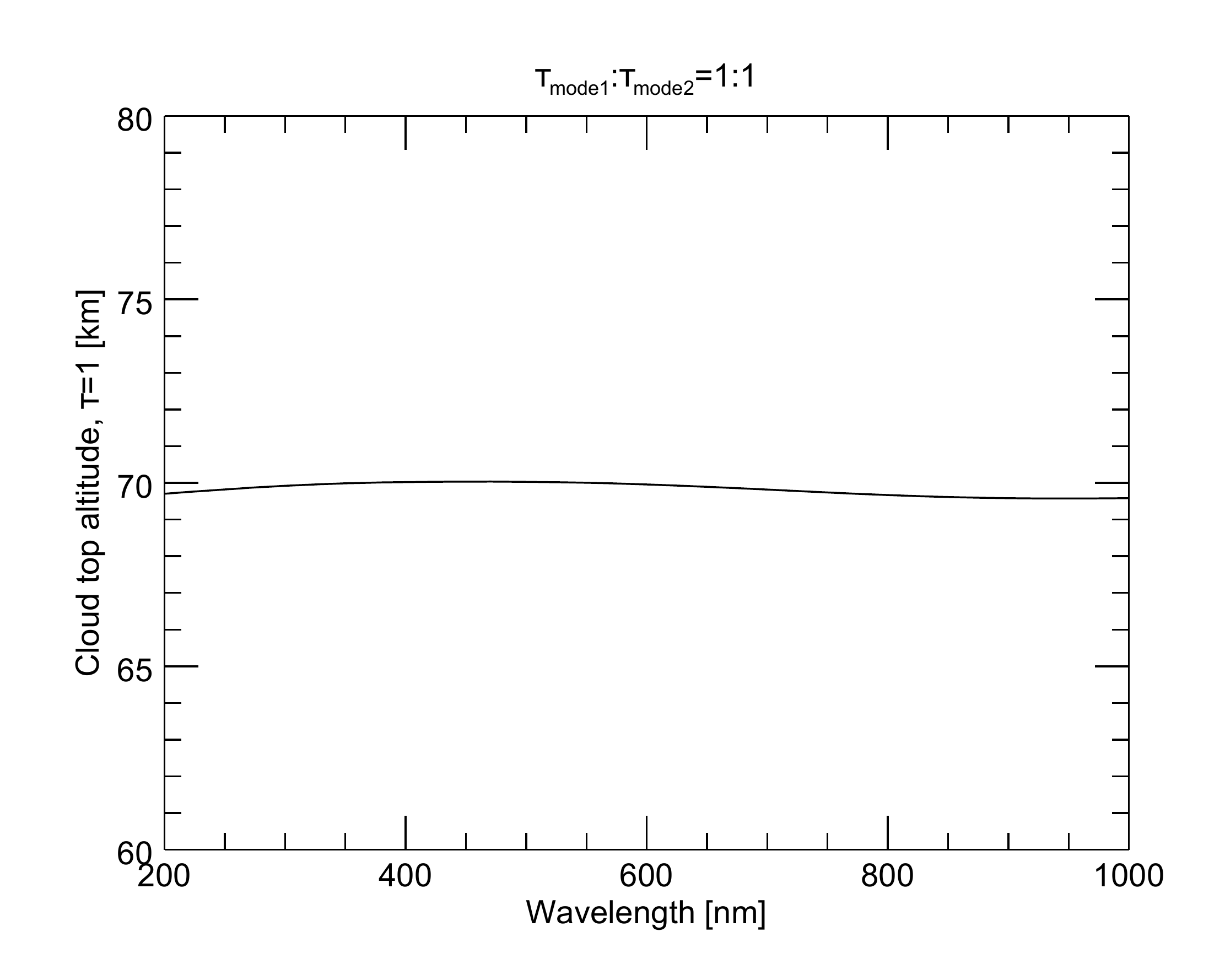}{0.5\textwidth}{(c)}
	}
	\caption{Spectral dependence of optical properties of the clouds. (a) Extinction cross-section. (b) Asymmetry factor, $g$. (c) Cloud top altitude (unity tau). \label{fig:atm_cloud}}
\end{figure*}

Two different sizes of cloud aerosols were taken into account in this study, the so-called `mode~1 and 2': mode~1 with r$_{\rm{eff}}$=0.43~$\mu$m and $\nu_{\rm{eff}}$=0.52 \citep{Pollack80b} and mode~2 with r$_{\rm{eff}}$=1.26~$\mu$m and $\nu_{\rm{eff}}$=0.076 \citep{Lee17} (Table~\ref{tab:cloud_aerosol}). The spectral dependence of the optical properties of the aerosols were calculated with the log-normal size distribution using a Lorentz-Mie code \citep{BOOK_Mishchenko02}. The refractive indices of the aerosols were taken from \citet{Hummel88} for 75\% H$_2$SO$_4$-H$_2$O aerosols. As shown in Fig.~\ref{fig:atm_cloud}a, the extinction coefficient of the aerosols shows little change over the spectral range of this study. The asymmetry factor ($g$) shows spectral variation (Fig.~\ref{fig:atm_cloud}b). We prescribed the relative abundance of modes 1 and 2 by imposing that the ratio of extinction between the two modes of aerosols are 1:1 at 365~nm. This extinction ratio is similar to the ratio in the upper cloud layer inferred from the Pioneer Venus Sounder Probe \citep{Pollack80b, Knollenberg80}. Our initial cloud top altitude was assumed to be 70~km at 365~nm that changes vertically by a scale height of 4~km, close to the value retrieved previously \citep{Ignatiev09,Lee12,Satoh15}. The cloud top altitude ($\tau=1$) shows little change over the entire wavelength range of this study (Fig.~\ref{fig:atm_gases}c). We later change this cloud top altitude to 64~km in Sect.~\ref{sec:results} to fit our observed I band reflectivity. This cloud top altitude is yet tentative, and we plan further studies to examine reliable cloud top altitude retrieval using the ground-based observations.

At first we simulated the reflectivity of Venus without the unknown absorber, which could not fit the observed reflectivity (Sect.~\ref{subsec:result_spectrum}). Later we took into account the unknown absorber to explain the observations (Sect.~\ref{subsec:result_UA}). We assumed that the unknown absorber is present in a 6~km thick layer whose middle altitude is located 3~km below the cloud top level, which is one of the best solutions in our previous study \citep{Lee21}. To simulate the unknown absorber, we reduced the single scattering albedo (SSA) of cloud aerosols within the 6~km layer by increasing the absorption relative to the extinction coefficient strictly attributable to the aerosols (R$_{\rm{UA}}$): SSA$=1-$R$_{\rm{UA}}$ \citep{Lee21}. This parameterization imitates the impact of a variable imaginary refractive index ($n_i$) of the cloud aerosols at our target wavelengths, since the SSA is the factor that is most sensitive to a variable $n_i$ amongst all the optical properties of the cloud aerosols, such as extinction cross section and $g$. This simplified approach helps us quantify the influence of the unknown absorber, even though we do not know its identity, for example, which size of aerosols would contain the absorber or, whether the absorber is solid or gaseous. In this study, we reduced the SSA of both mode~1 and 2 aerosols to account for the contribution of the unknown absorber. This assumption keeps the model configuration simple, and is consistent with the fact that all modes may contain the absorber as an impurity in particles \citep{Pollack80b}.

\begin{deluxetable*}{ccccc}
	\tablecaption{Size distributions of the cloud aeorols \label{tab:cloud_aerosol}}
	\tablewidth{0pt}
	\tablehead{\\
		& & & \multicolumn{2}{c}{Log-normal size distribution} \\
		\cline{4-5}
		\colhead{Mode} & \colhead{r$_{\rm{eff}}$ [$\mu$m]} &
		\colhead{$\nu_{\rm{eff}}$} & \colhead{$\bar{r}$  [$\mu$m]} & \colhead{$\sigma$}
	}
	\startdata
	mode 1 & 0.43 & 0.52 & 0.15 & 1.91 \\
	mode 2 & 1.26 & 0.076 & 1.05 & 1.31 \\
	\enddata
\end{deluxetable*}

\subsection{Radiative transfer model calculations} \label{subsec:RTM}
We used a Preconditioned Backward Monte Carlo (PBMC) algorithm \citep{GarciaMunoz15,GarciamunozMills2015} to fit the observed spectral reflectivity of the Venus disk during the campaign period. Solar phase angle ($\alpha$) was fixed to 80$^{\circ}$, and we calculated the multiple scattering processes in the atmosphere over the 180$-$1000~nm spectral range. We selected $d\lambda$=1~nm, which balances spectral resolution against computing efficiency. At each wavelength grid, we used 10$^6$ photons. The statistic error is 0.05\% over the spectral range. The gaseous absorption data set described in Sect.~\ref{subsec:atm} was convolved using a Gaussian function (FWHM=1~nm) to take into account fine absorption features (Fig.~\ref{fig:atm_gases}a and b). We confirmed that two results are consistent: one with the current configuration, and another that ran PBMC calculations at higher spectral resolution (0.005~nm) and then convolved into the 1~nm grid.

\section{Results} \label{sec:results}
\subsection{Temporal variation of the reflectivity}\label{subsec:result_obs_timeseries}
\begin{figure}
	\plotone{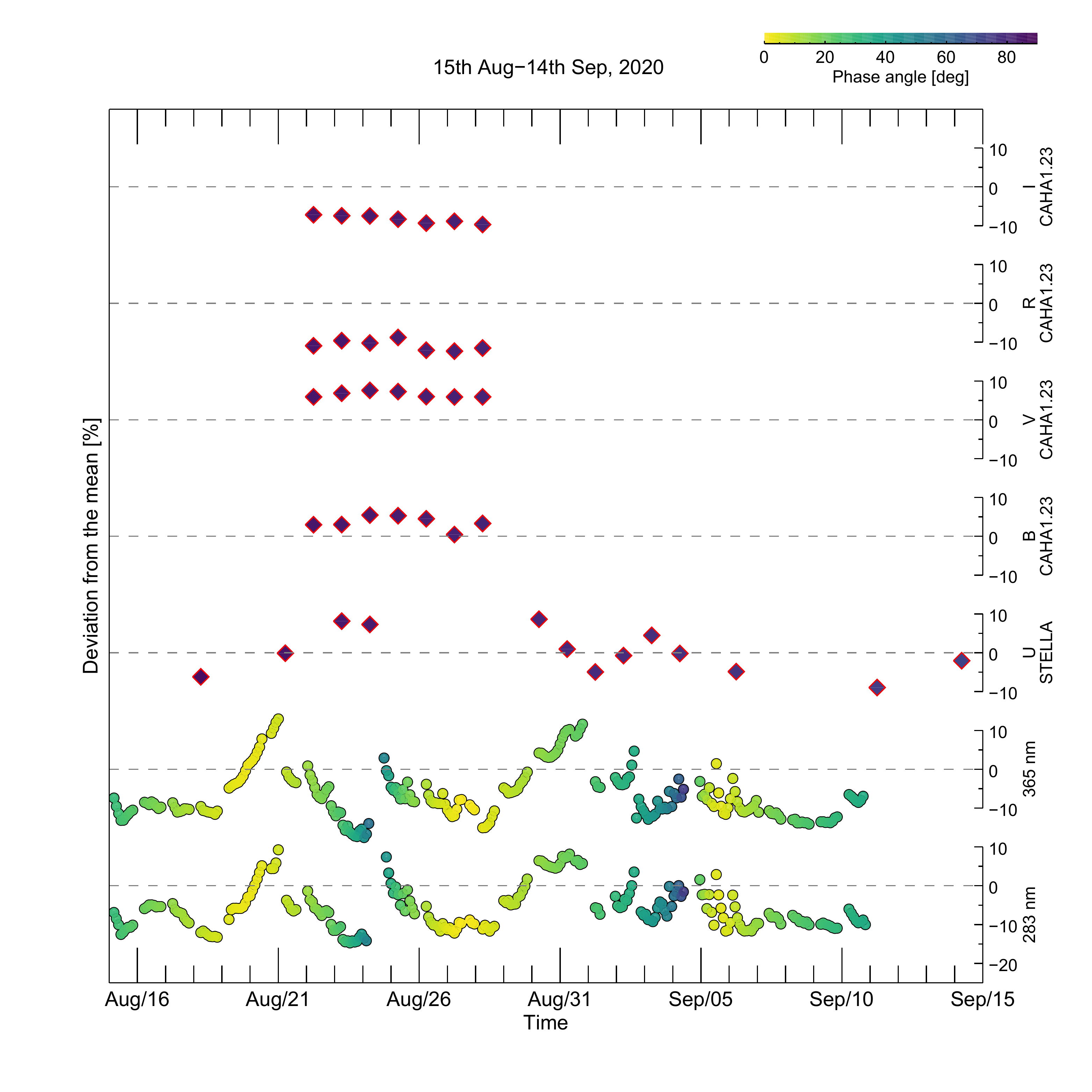}
	\caption{Time series of relative disk-integrated reflectivity [\%] compared to the mean (or reference) phase curves at each wavelength. Solar phase angle is shown as the color of symbols (colorbar, 0$-$90$^{\circ}$). The 283 and 365 nm data were taken by Akatsuki/UVI, which provided good temporal coverage. \label{fig:R_timeseries}}
\end{figure}

While the ground-based observations were mostly taken at $\alpha$$\sim$80$^{\circ}$ over the morning side (Fig.~\ref{fig:campaign_geometry}b), Akatsuki's viewing geometry changed along its orbit: between August 15 and September 15, $\alpha$ varied from 0$^{\circ}$ (full-Moon shape) to 74$^{\circ}$ (afternoon side). To compare the data taken from different viewing geometries, we calculated the relative brightness [\%], indicating how far the data at a specific time deviates from the reference phase curve $\overline{A_{\rm{disk-int}}}$ at the same phase angle (observed mean phase curves or a reference phase curve, such as \cite{Irvine68} and \cite{Mallama17}, as explained in Sect.~\ref{subsec:uvi}, \ref{subsec:caha1.23}, and \ref{subsec:stella}),
\begin{equation}\label{eq:rel_A}
 \frac{\left(A_{\rm{disk-int}}(\alpha,t)-\overline{A_{\rm{disk-int}}(\alpha)}\right)}{\overline{A_{\rm{disk-int}}(\alpha)}}\times100.
\end{equation}
The spatial distribution and absolute abundance of the unknown absorber are known to be variable over time \citep{Esposito80,DelGenio90,Markiewicz07a,Titov12,Lee19,Lee20} and this is also shown in Fig.~\ref{fig:R_timeseries} over a period of one month. The variation range of relative brightness at 365~nm reaches over 20\% (peak-to-peak). That at 283~nm shows a similar level of variation. The ground-based U band measurements (effective central wavelength at 365.6~nm) show also a considerable level of variation, implying a potential role of the ground-based U band imaging to track the temporal variations of the brightness. If its sampling frequency can be improved, e.g., by using more telescopes located in different longitudes, it should be also possible to resolve the 4$-$5~days periodicity of the brightness \citep{DelGenio82,Lee20}. The ground-based U band brightness shows about a day or two ahead temporal variations, e.g., the local brightness minimum on September 1st and the local peak on 3--4th, while the local min of UVI data is on 3rd and the local peak on 4--5th. This is consistent with the different viewing geometries: the morning side was observed by the ground-based U imaging and the noon-to-afternoon side by UVI (Fig.~\ref{fig:campaign_geometry}b). It will take $\sim$1$-$2~days for an air parcel to be drifted from morning to afternoon sides by the super-rotating background winds \citep{Sanchez-Lavega17,Horinouchi18}. Relative brightness at the B band shows a hint of temporal variations on August 22$-$28, especially the local minima on 23rd and 27th. This may be a day ahead variation compared to that of UVI, and caused by the unknown absorber, whose absorption extends towards the visible wavelength (Sect.~\ref{subsec:result_UA}). Future monitoring at the B band will be useful to quantify its short-term variations. At the VRI bands we do not expect to detect a short-term variability, because the impact of the unknown absorber diminishes at longer wavelengths (Sect.~\ref{subsec:result_UA}). A difference between B and V magnitudes can be a proxy to monitor the unknown absorber.

\subsection{Mean spectral reflectivity and comparison with model calculations}\label{subsec:result_spectrum}
The mean reflectivity spectrum of the campaign is shown in Fig.~\ref{fig:R_comparison}. All imaging data are corrected into that at the 80$^{\circ}$ solar phase angle, which is the value of ground-based observations near the end of August (Fig.~\ref{fig:campaign_geometry}b). This correction is done using a relative brightness [\%] at the time of observations and $\overline{A_{\rm{disk-int}}(80^{\circ})}$ (Eq.~\ref{eq:rel_A}). STELLA data is a mean over a relatively long period to compensate its yet undefined phase curve (Sect.~\ref{subsec:stella}). The normalized reflectivity of PMAS data at 445~nm (Sect.~\ref{subsec:pmas}) is adjusted to match the B band reflectivity (Sect.~\ref{subsec:caha1.23}). The PMAS data do not require any solar phase angle correction because the data were acquired at $\alpha$$\sim$80$^{\circ}$. The horizontal error bars indicate the bandwidths of imaging filters, and the vertical error bars the standard deviations over specific periods for each measurement as indicated in the legends.

\begin{figure*}
	\gridline{
		\fig{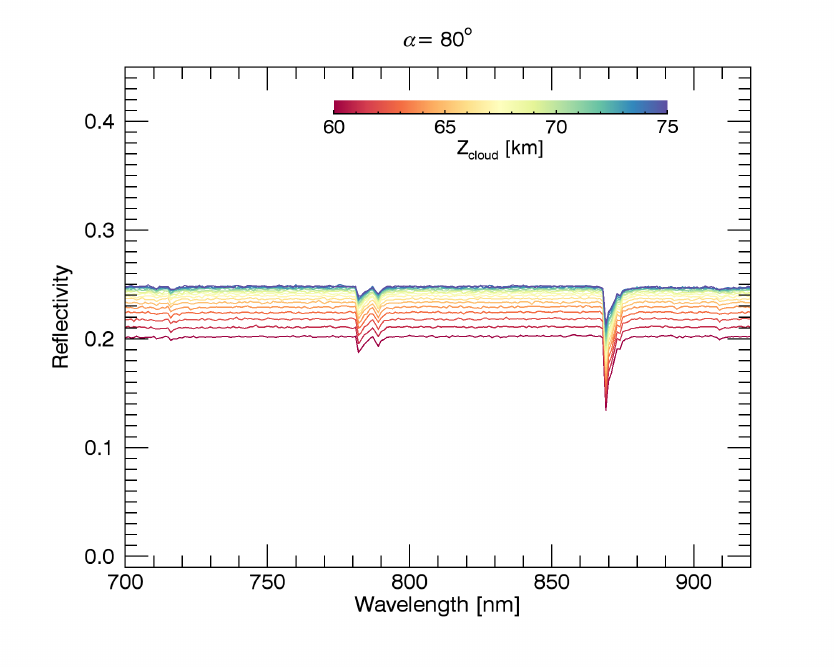}{0.5\textwidth}{(a)}
		\fig{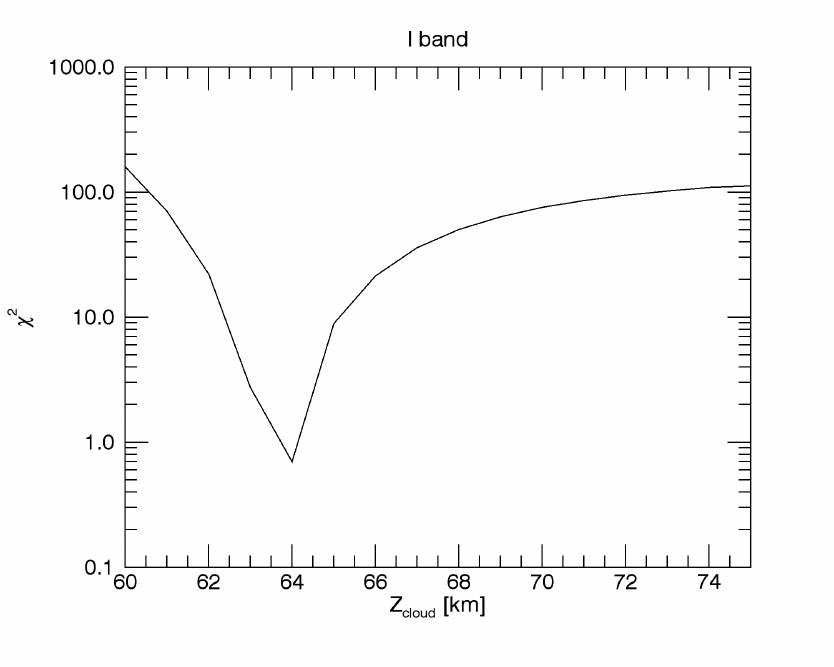}{0.5\textwidth}{(b)}
	}
\caption{Relation between the reflectivity over the wavelengths of the I band and cloud top altitude ($Z_{\rm{cloud}}$). (a) Variations of the calculated reflectivity according to $Z_{\rm{cloud}}$ that is changed between 60 and 75~km. (b) $\chi^2$ to fit the observed I band reflectivity as a function of $Z_{\rm{cloud}}$. \label{fig:cloud_top}}
\end{figure*}

The modeled reflectivity at $\alpha$=80$^{\circ}$ is compared in Fig.~\ref{fig:R_comparison} with two cloud top altitude assumptions, but without the unknown absorber. The large difference in reflectivity between the calculations and the observations are present in the wavelengths less than 460~nm. This difference is expected, as there is significant influence of the unknown absorber \citep{Perezhoyos18}. The calculated reflectivity with the 70~km cloud top shows slightly brighter than the observed VRI reflectivity. We compared the reflectivity at the I band with cloud top altitudes in the range between 60 and 75~km, which control the impact of Rayleigh scattering above the clouds (Fig.~\ref{fig:cloud_top}). We find that the 64~km cloud top altitude fits best the observed I reflectivity, and we use this cloud top altitude in further modeling calculations. But we note that this configuration of the cloud top altitude (64~km) should be considered as a tentative value, until we confirm the phase angle dependence using further observations.% (Sects.~\ref{subsec:caha1.23}).

The simulated reflectivity in Fig.~\ref{fig:R_comparison} shows that different cloud top altitudes can alter significantly the depth of SO$_2$ absorption at 283~nm. Under the assumed SO$_2$ abundance of 22.4~ppbv at 70~km \citep{Lee21}, we can compare the two cloud top altitude cases: 70~km and 64~km. 1) 70~km: the simulated spectrum results in shallow SO$_2$ absorption. The difference between the observed 283~nm reflectivity and the model suggests considerable absorption by the unknown absorber. 2) 64~km: the simulated spectrum presents significant SO$_2$ absorption. The unknown absorber's contribution may not be necessary to explain the 283~nm reflectivity. We conclude that to retrieve SO$_2$ gas abundance and absorption by the unknown absorber, it is necessary to have both a good UV spectral coverage over the SO$_2$ band and a reliable retrieval of the cloud top altitudes. Such an analysis was successfully done using UV spectrometer measurements by \citet{Marcq19}, although the retrieval of the cloud top altitude was somewhat limited. More accurate retrievals are planned for a future mission, VenSpec-U onboard EnVision \citep{Marcq21}. Also, future PHEBUS observations will contribute significantly to the determination of the SO$_2$ gas abundance by measuring the UV spectrum over the SO$_2$ band.

The reflectivity around 365~nm shows a difference between measurements by UVI and STELLA (Fig.~\ref{fig:R_comparison}). The reason of the difference is not clearly understood, but may be due to possible absolute calibration issue of UVI (Sect.~\ref{subsec:uvi}) and retrieval error for the STELLA analysis associated with the absence of the mean U band phase curve (Sect.~\ref{subsec:stella}). In the future, we will perform continuous star calibrations of UVI and Venus monitoring by STELLA to establish the mean U band phase curve.

\subsection{Unknown absorber}\label{subsec:result_UA}
\begin{figure}
	\plotone{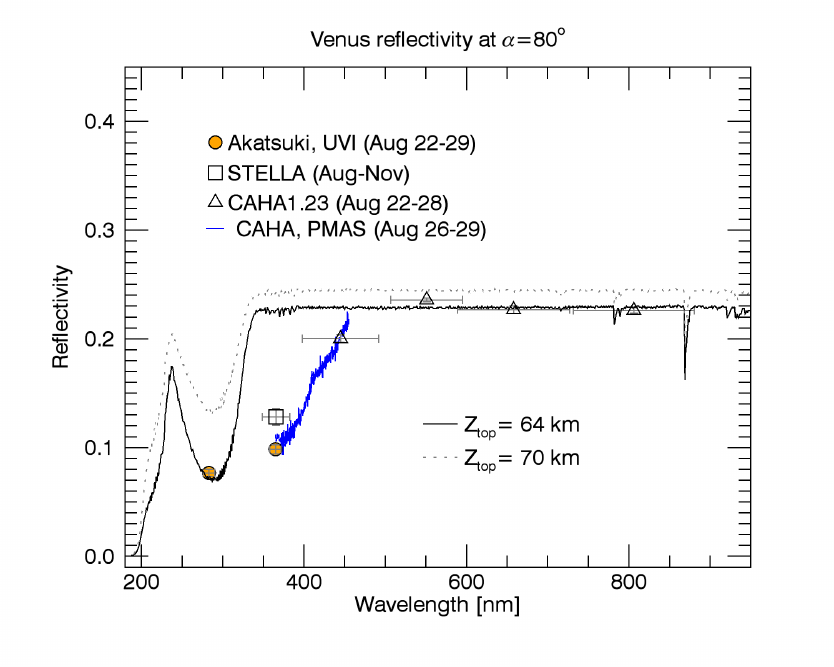}
	\caption{Comparison of Venus reflectivity obtained from the four facilities during the campaign: Akatsuki/UVI (Sect.~\ref{subsec:uvi}), STELLA/WiFSIP (Sect.~\ref{subsec:stella}), CAHA1.23/DLR-MKIII (Sect.~\ref{subsec:caha1.23}), and CAHA3.5/PMAS (Sect.~\ref{subsec:pmas}). The vertical error bars indicate standard deviations, and the horizontal error bars mean the band width (FWHM). The calculated reflectivity is compared together (Sect.~\ref{sec:model}), which assumed two cloud top altitudes without the unknown absorber (solid and dashed lines). \label{fig:R_comparison}}
\end{figure}

\begin{figure}
	\plotone{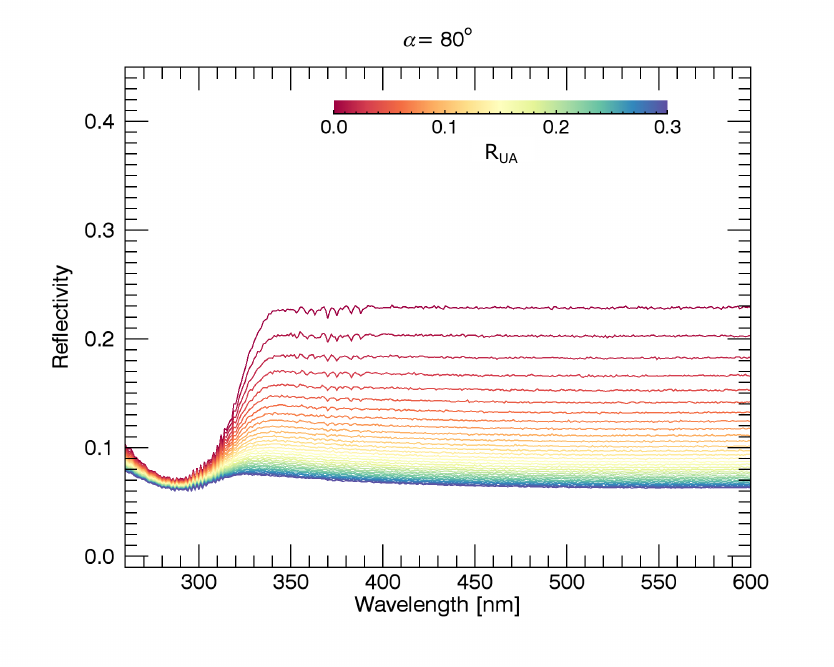}
	\caption{Simulated reflectivity at $\alpha$=80$^{\circ}$ with a range of R$_{\rm{UA}}$ from 0.0 to 0.3 (see Sect.~\ref{subsec:clouds}). This assumes the best fit of the cloud top in Fig.~\ref{fig:cloud_top}b (64~km). The unknown absorber is assumed to be within the 6~km thickness layer whose middle is located 3~km below the cloud top level \citep{Lee21}, i.e., 58$-$64~km. Within this layer, SSA of the cloud aerosols is reduced by R$_{\rm{UA}}$. \label{fig:Abs_ratio}}
\end{figure}

Figure~\ref{fig:R_comparison} shows the difference in reflectivity between the simulations and the observations over the $\sim$350$-$460~nm range. This difference is due to the absorption by the unknown absorber in the clouds, which was not included in the simulations. In this section, we estimate the contribution of the unknown absorber using the PMAS spectral data, which provide the wavelength coverage down to the atmospheric cutoff in the UV. Our assumption on the unknown absorber is described in Sect.~\ref{subsec:clouds}: reducing SSA of cloud aerosols by increasing R$_{\rm{UA}}$ within the 6-km layer right below the cloud top. The simulated reflectivity is wavelength dependent due to gaseous absorption and increasing Rayleigh scattering towards short wavelengths as shown in Fig.~\ref{fig:Abs_ratio}. We prepared a table of expected reflectivity as a function of wavelength (d$\lambda$=1~nm) and R$_{\rm{UA}}$ (dR$_{\rm{UA}}$=0.01). Then we compared this table and the observed PMAS reflectivity (Fig.~\ref{fig:R_comparison}) to find corresponding absorption along wavelengths. This process was done in a 2-dimensional interpolation: linear interpolation along wavelengths and least squares quadratic interpolation for the absorption at a fixed wavelength to take into account considerable non-linear property with R$_{\rm{UA}}$, as shown in Fig.~\ref{fig:Abs_ratio}.

\begin{figure}
	\plotone{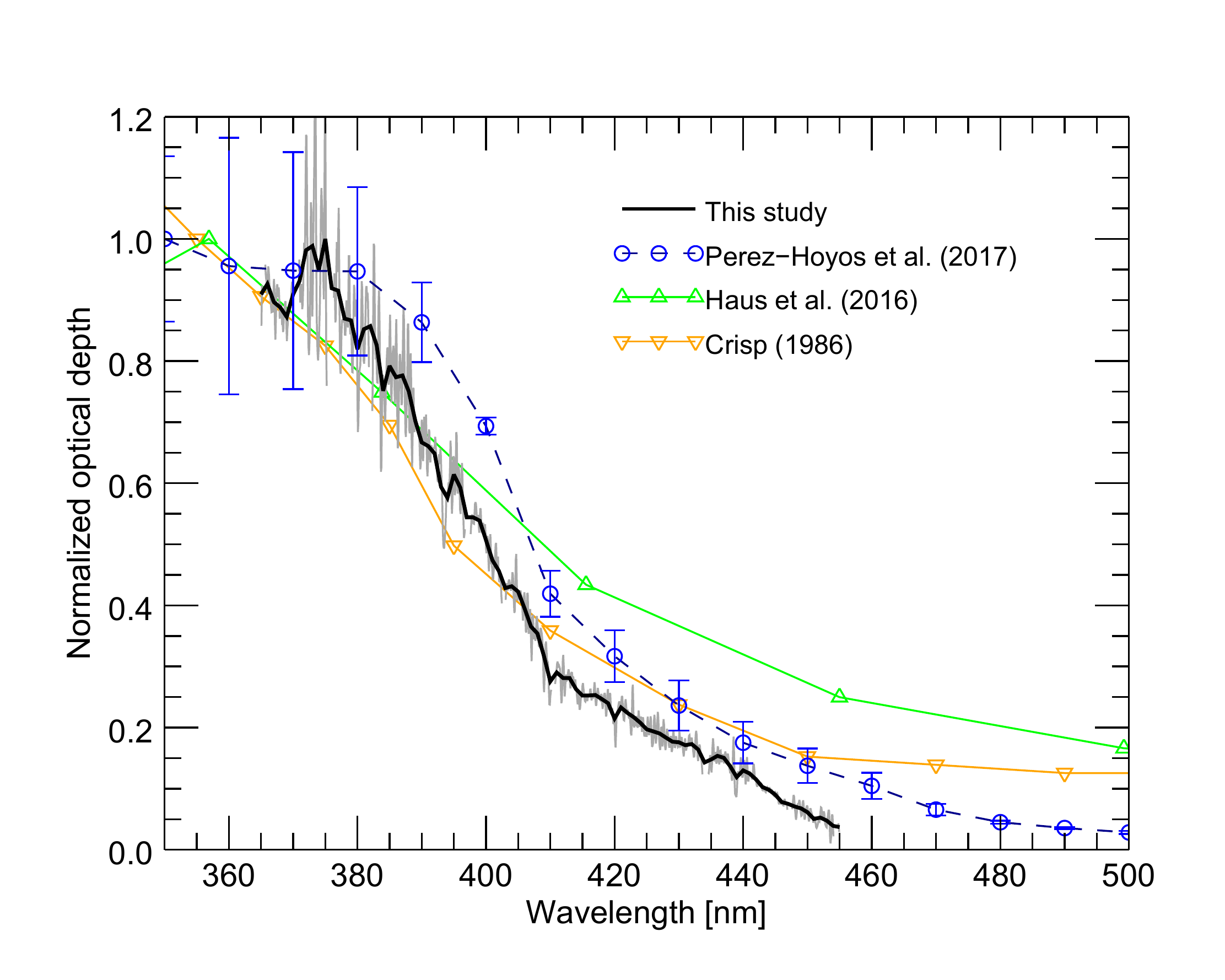}
	\caption{Relative optical depth of the unknown absorber (normalized to the maximum in the 350$-$500~nm range). Required absorption to match the PMAS data is shown with the gray line (Sect.~\ref{subsec:result_UA}), and its convolution (FWHM=1~nm) is shown with the black line. Previous observational data analysis \citep{Perezhoyos18} is shown for comparison (blue circles). Assumptions that were applied for the solar heating rate calculations (up/down triangles) \citep{Crisp86, Haus16} are shown together. \label{fig:Optdepth}}
\end{figure}

The resulting absorption spectrum is shown in Fig.~\ref{fig:Optdepth}, as normalized optical depth to the maximum value in the 350$-$500~nm range. So the possible errors related with the cloud top altitude (Sect.~\ref{subsec:result_spectrum}) become irrelevant, and we can focus on the spectral shape of the absorption to compare with those in previous studies \citep{Crisp86,Haus16,Perezhoyos18}. The high spectral resolution (gray curve) are convolved into the 1~nm grid (black curve), which is our model calculation resolution. The relative optical depth decreases with increasing wavelengths, but more rapidly at $\lambda$$<$410~nm than that at longer wavelengths, and expected to close to zero at the V band. This wavelength dependence is consistent with the relative optical depth at low latitudes \citep{Perezhoyos18}. This is an expected result, considering the large portion of low latitudes in the equatorial view from the Earth (Fig.~\ref{fig:campaign_geometry}b). This consistent spectral dependency also implies a negligible change in chemical composition of the unknown absorber between the afternoon equatorial region in 2007 and the morning side in 2020.

Compared to the assumptions that were used in the solar heating rate calculations in \citet{Crisp86} and \citet{Haus16}, both observational data analyses suggest weaker absorption in the visible wavelength range than those in the UV. This spectral shape of absorption may alter the solar heating rate near the cloud top level atmosphere. However, there are more factors that can affect the solar heating, such as the unknown absorber's absolute abundance \citep{Lee19} and vertical location \citep{Crisp86,Haus16,Lee21}, and the cloud top vertical structure \citep{Lee15b}. Updating the heating rate is not within the scope of this study, but may be possible in future studies.

\section{Summary and perspective of future campaigns} \label{sec:discussion}

Our dayside observation campaign was conducted with the PHEBUS spectrometer on board BepiColombo and the UVI camera on board Akatsuki to better understand the UV absorbers in the Venusian clouds. Our campaign was designed to cover a broad wavelength range from 52 to 1700~nm thanks to Earth-bound observation facilities. Despite the fact that our data analysis eventually could only utilize the data between 283 and 800~nm wavelength (Sect.~\ref{sec:obs}), we achieved the following goals and insights:

\begin{enumerate}
	\item We successfully accomplished the Venus observation campaign using multiple ground- and space-based facilities almost simultaneously.
	
	\item Despite the challenging brightness of the target (too bright), we managed to acquire high quality data.
	
	\item The PHEBUS team could establish a robust observation strategy to make successful Venus observations in future opportunities, e.g., in June and July 2022.
	
	\item Using the campaign data in the 283$-$800~nm range we retrieved the relative optical depth of the unknown absorber on the morning side disk. Our result is consistent with the previous report using the data acquired in 2007 over the afternoon equatorial region \citep{Perezhoyos18}.
	
	\item We plan future campaigns to retrieve both SO$_2$ gas abundance and the absorption by the unknown absorber in the 180$-$450~nm range, using data acquired by PHEBUS (180$-$320~nm), UVI (283 and 365~nm), and ground-based telescopes (350-800~nm).
	
	\item We established that flux measurements at the VRI bands can provide a constraint on the cloud configuration to generate simulated reflectivity. We will continue VRI imaging in future campaigns.

	\item The U band phase curve of Venus is poorly defined. We plan to continue U band imaging to define a mean phase curve.
	
	\item Through the ground-based U band measurements may be possible to track the temporal variability of Venus' reflectivity, in addition to space-based measurements.
		
	\item Akatsuki's UV imaging is an excellent reference to compare short-term variations.
	
	\item PMAS observation and flux measurements at the B band will be repeated in our future campaigns to understand possible temporal variations of the unknown absorber.
\end{enumerate}

\section*{Acknowledgements}
This research used the data collected at the Centro Astron\'{o}mico Hispano-Alem\'{a}n (CAHA) at Calar Alto, operated jointly by Junta de Andaluc\'{i}a and Consejo Superior de Investigaciones Cient\'{i}ficas (IAA-CSIC). This research has made use of the integral-field spectroscopy data-reduction tool p3d, which is provided by the Leibniz-Institut für Astrophysik Potsdam (AIP). Akatsuki/UVI data are publicly available at the JAXA archive website, DARTS (\url{http://darts.isas.jaxa.jp/}), and the NASA archive website, PDS (\url{https://pds.nasa.gov/}). UVI level 3x products (L3bx) were used in this study \citep{Murakami_dataset}. This study used the TSIS-1 SIM data (Version 06, doi:\url{https://doi.org/10.25810/y9f8-ff85}). MK and OE thank TUBITAK National Observatory for a partial support in using T100 telescope with project number 20CT100-1688. RH and ASL have been supported by the Spanish project PID2019-109467GB-I00 (MINECO/FEDER, UE) and Grupos Gobierno Vasco IT-1366-19. PK and MS acknowledge support from grant LTT-20015.

\appendix
\section{Spectral signature of methane (CH$_4$)}\label{Appendix}
At first we assumed possible methane gas in the model calculations (Sect.~\ref{subsec:atm}). But later we found that its spectral signature should be clear to detect with remote observations. We excluded methane for the results in this manuscript (Sect.~\ref{sec:results}). Confirmation of possible methane may be a subject of future observation projects. 
\begin{figure}
	\plotone{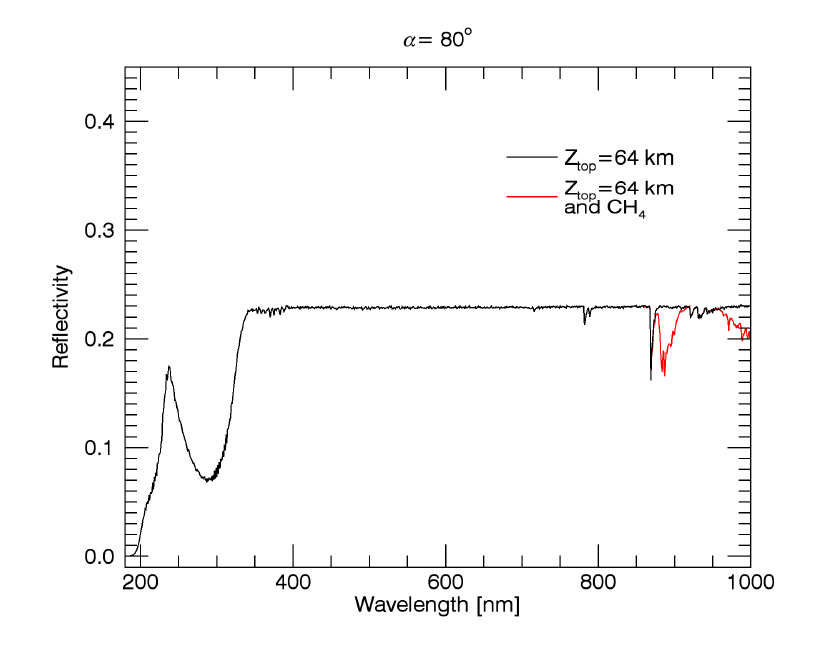}
	\caption{Simulated Venus reflectivity for two cases: without CH$_4$ (black) and with CH$_4$ (red). The latter case assumed the vertical mixing ratio of methane in Fig.~\ref{fig:atm_gases}c. \label{fig:CH4}}
\end{figure}

\newpage

\bibliography{paper}{}
\bibliographystyle{aasjournal}

\end{document}